\documentclass{aa} 
\usepackage{graphicx}
 \usepackage{aalongtable}
\usepackage{txfonts}
\usepackage{natbib}
\def\Teff{$\rm T_{eff}$}
\def\grav{$\log g$}
\def\MuLeo{$\mu$Leo}
\def\VT{$\xi$}

\def\GF{$\log\, gf$}

\catcode`\@=11
\def\gsim{\ifmmode{\mathrel{\mathpalette\@versim>}}
    \else{$\mathrel{\mathpalette\@versim>}$}\fi}
\def\lsim{\ifmmode{\mathrel{\mathpalette\@versim<}}
    \else{$\mathrel{\mathpalette\@versim<}$}\fi}
\def\@versim#1#2{\lower 2.9truept \vbox{\baselineskip 0pt \lineskip 
    0.5truept \ialign{$\m@th#1\hfil##\hfil$\crcr#2\crcr\sim\crcr}}}
\catcode`\@=12
\begin{document}
   \title{The metallicity distribution of bulge clump giants in Baade's Window
\thanks{Based on ESO-VLT 
observations during Paris Observatory FLAMES GTO
71.B-0196.}$^{,}$\thanks{Full Tables 1, 2 and 3 are only available in electronic form
  at http://www.andaa.org}
}

   \author{V. Hill\inst{1,2}
	  \and
	  A. Lecureur\inst{2}
          \and
	  A. G\'{o}mez\inst{2} 
	  \and 
	  M. Zoccali\inst{4}
	  \and 
	  M. Schultheis\inst{3}
	  \and 
	  C. Babusiaux\inst{2} 
	  \and 
	  F. Royer\inst{2}
	  \and 
	  B. Barbuy\inst{5}
	  \and 
	  F. Arenou\inst{2}
	  \and 
	  D. Minniti\inst{4}
	  \and 
	  S. Ortolani\inst{7}
          }

   \offprints{V. Hill}

   \institute{Universit\'e de Nice Sophia Antipolis, CNRS, Observatoire de la
  C\^ote d'Azur, bd. de l'Observatoire, B.P. 4229, 06304 Nice Cedex 4, France
             \email{Vanessa.Hill@oca.eu}
	 \and
	 GEPI, Observatoire de Paris, CNRS UMR 8111, Universit\'e Paris Diderot ; Place Jules Janssen 92190 Meudon, France
              \email{Aurelie.Lecureur@obspm.fr,
  Carine.Babusiaux@obspm.fr, Ana.Gomez@obspm.fr,
  Frederic.Royer@obspm.fr, Frederic.Arenou@obspm.fr}  
      	 \and
	 Observatoire de Besan\c{c}on, CNRS UMR 6091, BP 1615, 25010 Besan\c{c}on, France
	     \email{mathias.schultheis@obs-besancon.fr}
	 \and
             P. Universidad Cat\'olica de Chile, Departamento de 
	     Astronom\'\i a y Astrof\'\i sica, Casilla 306, Santiago 22, 
	     Chile \email{mzoccali@astro.puc.cl, dante@astro.puc.cl}
	 \and
	     Universidade de S\~ao Paulo, IAG, Rua do Mat\~ao 1226,
             S\~ao Paulo 05508-900, Brazil
	     \email{barbuy@astro.iag.usp.br}
	 \and
	     Universita di Padova,Vicolo dell'Osservatorio 5, I-35122 Padova, 
	     Italy	
	     \email{sergio.ortolani@unipd.it}
             }

   \date{Received; accepted }

  \abstract
   {}
{We seek to constrain the formation of the Galactic bulge by means of
  analysing the detailed chemical composition of a large sample of red
  clump stars in Baade's window. 
}
   {We measure [Fe/H] in a sample of 219 bulge red clump stars
  from R=20000 resolution spectra obtained with FLAMES/GIRAFFE at the
  VLT, using an automatic procedure, differentially to the metal-rich
  local reference star \MuLeo. For a subsample of 162 stars, we also
  derive [Mg/H] from spectral synthesis around the
  \ion{Mg}{I} triplet at $\rm\lambda$ 6319 \AA.  
}
{The Fe and Mg metallicity distributions are both asymmetric, with median
values of $+0.16$ and $+0.21$ respectively. The iron distribution is clearly
  bimodal, as revealed both by a deconvolution (from observational
  errors) and a Gaussian decomposition. 
The decomposition of the observed Fe and Mg metallicity distributions into Gaussian
components yields two populations of equal sizes
(50\% each): a metal-poor component centred around [Fe/H]~$\rm=-0.30$
and [Mg/H]~$\rm=-0.06$ with a large dispersion and a narrow metal-rich
component centred around $\rm [Fe/H]=+0.32$ and $\rm [Mg/H]=+0.35$. The metal poor
component shows high [Mg/Fe] ratios (around 0.3) whereas stars in the
metal rich component are found to have near solar
  ratios. Babusiaux et al. (2010) also find kinematical differences between the two
components: the metal poor component shows kinematics compatible with
an old spheroid whereas the metal rich component is consistent with a
population supporting a bar. In view of their chemical and kinematical
properties, we suggest different formation scenarii for the two
populations: a rapid formation timescale as an old spheroid for the
metal poor component (old bulge) and for the metal rich component, a
formation over a longer time scale driven by the evolution of the bar
({\em pseudo-bulge}). 
}
{
Guided by these results, we build a simple model combining two
components: a simple closed box model to predict the metal
poor population contribution, whereas the metal rich 
population is modelled using the observed local thin disc metallicity
distribution, shifted in metallicity. The {{\em pseudo-bulge}} is compatible with being formed from the inner thin
disc, assuming large (but plausible) values of the gradients in the early
Galactic disc. 
} 
   \keywords{Galaxy: bulge -- Galaxy: formation -- Galaxy: abundances --
             Stars: abundances -- Stars: atmosphere
	      }

   \maketitle
%

\section{Introduction}

The Milky-Way bulge has been the subject of quite intense debates in
the community, as its status is not yet fully established. With
various stellar population characteristics similar to those of the central old spheroids
found at the centre of earlier-type galaxies, and others (mostly geometrical and kinematical) that rather
point towards a very strong influence of a bar responsible for a {\em
  pseudo-bulge}, it sits at the border between these two types of bulges.
The formation scenarios for bulges can be classified in three different types: 
(i) initial collapse of gas at early times \citep[see e.g. ][]{Eggen1962};
(ii) merging subclumps, either through an early disc evolution
\citep{Noguchi1999,Immeli2004}, or through mergers
\citep{Aguerri2001,Scannapieco2003,Nakasato2003}; 
(iii) secular evolution of the Galactic disc
\citep{Combes1981,Pfenniger1990,Raha1991}. The merger scenario itself
is similar to an early collapse scenario from the point of view of
the formation characteristic timescale (early and fast).
 It was also recently suggested that the Galactic bulge could be the
result of both formation processes, with an {\it old spheroid} complemented
by a {\em bar-driven pseudo-bulge}
\citep{Nakasato2003,Kormendy2004,Gerhard2006}. The relative importance of
the two processes (or even populations) however remains to be
established. 

 The presence of a bar in the inner Galaxy has been suggested by 
\citet{deVaucouleurs1964} from gas kinematics and confirmed since then by
 numerous studies including 
infrared surface brightness map \citep[e.g.][]{2MASS_survey},
star-counts, 
red-clump distances \citep[e.g.][]{Babusiaux2005,Nishiyama2006,Rattenbury2007a}, 
microlensing and stellar kinematics \citep[e.g.][]{Zhao1994,Howard2008,Howard2009}. 
The boxy aspect of the bulge, detected in the infrared
 light profile \citep[e.g.][]{Dwek1995}, also argues for a {\em
 pseudo-bulge} secularly evolved from the galactic disc.


On the other hand, photometric studies in selected windows on the Galactic
bulge, in the visible and the near infrared, soon made apparent
that the stellar populations in the bulge is old and
metal-rich \citep{Ortolani1995,Feltzing2000,Kuijken2002,Zoccali2003,Clarkson2008}. 
Spectroscopically, \citet{Rich1988} obtained one of the first
metallicity distribution in Baade's window based on low-resolution spectra
of M giants, followed by \citet{Ibata1995a,Ibata1995b} using K giants and \citet{Sadler1996} using red clump stars,
all finding a large metallicity dispersion. High resolution spectra of a
limited number of stars (10 to 20) in Baade's window
\citep{McWilliam1994,Fulbright2006,Fulbright2007}, and more 
recently by our group \citep{Zoccali2006,Lecureur2007} in a larger
sample of $~$50 stars in four windows of the Galactic bulge, showed
enhanced [$\alpha$/Fe], compatible with a fast chemical enrichment of
the Galactic bulge. Both the chemical and age properties of stellar
popultions in the Galactic bulge thus point towards a rapid bulge
formation. 
Early combined metallicity and kinematics measurements also pointed
towards bulge formation through dissipational collapse
\citep{Minniti1996}. 

More recently still, our group presented in \citet{Zoccali2008} the
first metallicity distribution entirely based on high-resolution
spectra of large sample of red giant branch stars in three fields of the Galactic bulge
(close to the minor axis, at b$\sim -4, -6$ and $-12^{\circ}$).
In this paper we will show how determining metallicity distributions
from a large ($> 200$) and almost uncontaminated sample of red clump
stars in Baade's window, observed at high spectral resolution can lead to significant
improvements in our understanding of the origin and nature of the
Galactic bulge. In particular, we show how using two independent
elements (iron {\em and} magnesium) as metallicity tracers (and their
ratio [Mg/Fe]) reveals the nature of the stellar population. 
In Sect.~2, we present our target selection and observations with
FLAMES on VLT, in Sect.~3 we detail the stellar parameter and elemental
abundance measurement methods, while Sect.~4 examines the issues of
sample representativity and contamination. Finally, in Sect.~5, we
discuss the resulting [Fe/H] and [Mg/H] metallicity distributions and
[Mg/Fe] trends, show that the sample can be separated into two
distinct populations and interpret these in the framework of various
formation scenarios and chemical evolution models. Sect.~6 gives our
conclusions from this work.

\section{Target selection, observations, and sample representativity}

\subsection{Target selection}

\begin{figure}
   \centering
   \resizebox{\hsize}{!}{\includegraphics{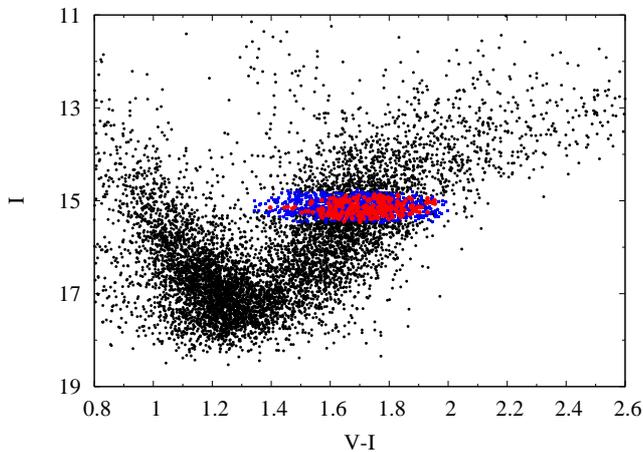}}
      \caption{(I,(V-I)) colour magnitude diagram for the Baade's
      window field BUL-SC45 from the OGLE II survey. The stars
      identified as red clump members by the OGLE survey are indicated
      in blue and the targets observed with FLAMES/GIRAFFE are
      indicated in red.}  
         \label{CMD_BW}
\end{figure}

We selected a sample of stars in the  Baade's window field BUL-SC45
from the OGLE II survey \citep{Udalski1997,Paczynski1999}, among the
$\rm\sim$ 1400 stars identified as red clump members by the OGLE
survey (see Fig. \ref{CMD_BW}). The sample was first restricted to
targets also present in the 2MASS \citep{2MASS_survey} and DENIS
surveys \citep{DENIS_survey}, in order to expand the photometric
coverage to the IR, which  is an asset for photometric temperature
determination (see Sect.~\ref{Determine_stellar_param}). From this
subsample ($\rm\sim$ 800 stars), a further restriction was applied in
the ($\rm K_0$, $\rm (J-K)_0$) colour magnitude diagram to lower a
possible disc contamination: stars with $\rm K_0 < 12$ and $\rm
(J-K)_0 < 0.5$ were rejected, as they lie in a locus where dwarfs are
expected to extend. To do so, colours were dereddenned using the mean
extinction value for the Baade's window $\rm A_ V=1.5$
\citep{Paczynski1999} and the reddening law $\rm A_K/A_V=0.089$
\citep{Glass1999}. Stars with $\rm 16.5 < V < 17.2$ were then selected
for observations with FLAMES-GIRAFFE, letting the fibre allocation
procedure randomly pick the final selection of 228 stars (see
Fig. \ref{CMD_BW}). The sample OGLE and 2MASS identification and
photometry are given in Table~\ref{Table_photo}, available at the CDS
in its entirety (the first lines of the table are reproduced in the
printed version).
To optimise the exposure time, the total sample was divided in two
subsamples: (a) 114 stars with $\rm 16.5<V<16.9$ and (b) 114 stars
with $\rm 16.9<V<17.2$. In each subsample, $\sim$16 fibres were
devoted to blank sky regions to allow for proper sky-subtraction. In
total, 130 fibres were allocated in each of the two GIRAFFE fibre
configurations (i.e. close to the maximum allowed).   

\subsection{Observations}
The observations were performed during August 2003 with the
ESO/VLT/FLAMES facility, as part of the Guaranteed Time Observation
programs of the Paris Observatory (PI, A.G\'omez). The spectra were
obtained with the GIRAFFE spectrograph in Medusa mode using two
high-resolution settings: HR13 et HR14, yielding resolving powers $\rm
R=22500$ and 28800, respectively. Spectral coverage spans from 6120 to
6405 \AA\ for HR13 and from 6383 to 6626 \AA\ for HR14, and the total
exposure times achieved were of 2h15 and 3h00 in HR13 for sample a)
and b) respectively, and 4h30 and 6h00 in HR14 for sample a) and b)
respectively. 
Using the FLAMES link to the UVES spectrograph, a subsample of 12
stars were simultaneously observed at higher resolutions ($\rm
R\sim48000$). This subsample was previously described in
\cite{Zoccali2006} and \cite{Lecureur2007}.  

\begin{table*}[!htbp]
\begin{center}
\caption{Sample identifications, BVRI OGLE and JHK 2MASS
  photometry. This table is available electronically in its entirety,
  including also errors on the photometric bands}
\label{Table_photo}
\begin{tabular}{llrrrrlrrr}
\hline
\hline
\multicolumn{1}{c}{ID} & \multicolumn{1}{c}{OGLE-ID} & 
\multicolumn{1}{c}{B} & 
\multicolumn{1}{c}{V} & 
\multicolumn{1}{c}{R} & 
\multicolumn{1}{c}{I} & 
\multicolumn{1}{c}{2MASS-ID} &
\multicolumn{1}{c}{J} & 
\multicolumn{1}{c}{H} & 
\multicolumn{1}{c}{K}  \\
& & mag & mag & mag & mag & & mag & mag & mag \\
\hline 
BWc-1  & BUL-SC45 393125 & 20.985 & 18.587 & 16.838 & 15.010 & J18035033+3005324 & 13.647 & 12.857 & 12.677 \\
BWc-2  & BUL-SC45 545749 & 20.434 & 18.830 & 17.191 & 15.392 & J18035671+3005378 & 14.115 & 13.312 & 13.178 \\
BWc-3  & BUL-SC45 564840 & 20.179 & 18.689 & 16.906 & 15.144 & J18035461+3001064 & 13.845 & 13.087 & 12.720 \\
BWc-4  & BUL-SC45 564857 & 19.593 & 18.240 & 16.760 & 15.161 & J18035531+3000576 & 13.941 & 13.111 & 12.886 \\
BWc-5  & BUL-SC45 575542 & 21.116 & 18.750 & 16.982 & 15.175 & J18035592+2955439 & 13.909 & 13.267 & 13.030 \\
BWc-6  & BUL-SC45 575585 & 19.497 & 18.238 & 16.744 & 15.069 & J18035640+2955122 & 13.832 & 13.097 & 12.955 \\
BWc-7  & BUL-SC45 575585 & 19.497 & 18.238 & 16.744 & 15.069 & J18035640+2955122 & 13.832 & 13.097 & 12.955 \\
BWc-8  & BUL-SC45  78255 & 20.702 & 18.616 & 16.972 & 15.163 & J18031236+3003596 & 13.724 & 12.990 & 12.885 \\
BWc-9  & BUL-SC45  78271 & 20.588 & 18.527 & 16.903 & 15.131 & J18031656+3003517 & 13.885 & 13.153 & 12.950 \\
BWc-10 & BUL-SC45  89589 & 19.790 & 18.213 & 16.695 & 15.058 & J18031877+3001101 & 13.401 & 12.709 & 12.505 \\
...    &  ...            & ...    & ...    & ...    & ...    & ...               & ...    & ...    & ...\\
\hline
\end{tabular}                
\end{center}
\end{table*}

The data reduction was carried out using the girBLDRS \footnote{see
  http://girbldrs.sourceforge.net/} pipeline developed at the Geneva
Observatory \citep{Blecha2000} which includes cosmic-ray removal, bias
subtraction, flat-field correction, individual spectral extraction and
accurate wavelength calibration based on simultaneous calibration
exposures. Co-addition of the individual spectra and sky substraction
was performed independently from the girBLDRS, using several IRAF
tasks. For each setup, an ``average sky'' was made by combining the 16
sky fibre spectra and subtracted to the spectra of each target. All
the spectra for each star were then co-added with a $\kappa$-sigma clipping
to remove the cosmic rays.

\subsection{Contamination and sample representativity}\label{section_contam}

Since all RGB stars experience the core helium burning phase, red
clump stars are excellent tracers for the metallicity distribution of
a given stellar population. Moreover, as mentioned by
\citet{Fulbright2006}, the metallicity-dependent lifetime of the
horizontal branch (HB) has a negligible effect on the metallicity
distribution (MD). These authors estimated that an increase of 1 dex
corresponds to an increase of 10\% in the HB lifetime, leading to a
correction of -0.02/-0.03 on the MD. Red clump giants are also good
candidates to sample the bulge MD because the CMD region where they
stand is known to be little contaminated by the other Galactic
components \citep{Sadler1996, Fulbright2006}. 
However, since stellar evolution models predict that the less metallic
stars burn helium in their cores on the blue horizontal branch (and
not in the clump), it is on the other hand expected that there is a
minimum metallicity that red clump stars can sample. We checked how
our sample selection centred on the red clump would have missed metal
poor stars by the inspection of 9 and 12Gyr old isochrones by 
\citet{Girardi2000} with various metallicities, in J-K, V-I, 
V-K and V-K bands, and compared them with our selection box 
(allowing for reddening). We thereby estimated that stars down to
[Fe/H]$\rm\geqslant-1.3$ or even $-1.5$ could be included in the sample selection box.  
However, because mass-loss is poorly known and plays a major role in
shaping the horizontal branch, it is difficult to quantify, for each metallicity bin, the bias that is introduced by selecting the red clump. 
However, the fact that the MD decreases
smoothly up to our metal-poor limit suggests that we might be loosing
only a few stars. Furthermore, the comparison of our red clump based 
MD to that obtained by \citet{Zoccali2008} from a sample selected higher 
up on the RGB with no such built-in bias (see Sec. \ref{SecCompRGB-RC}), 
shows no significant difference between the two samples on the metal-poor 
side (that however only reaches to $\rm [Fe/H]\sim -1.2$ to $-1.5$\,dex).

\begin{figure}   
\centering   
\includegraphics[angle=0,width=8cm]{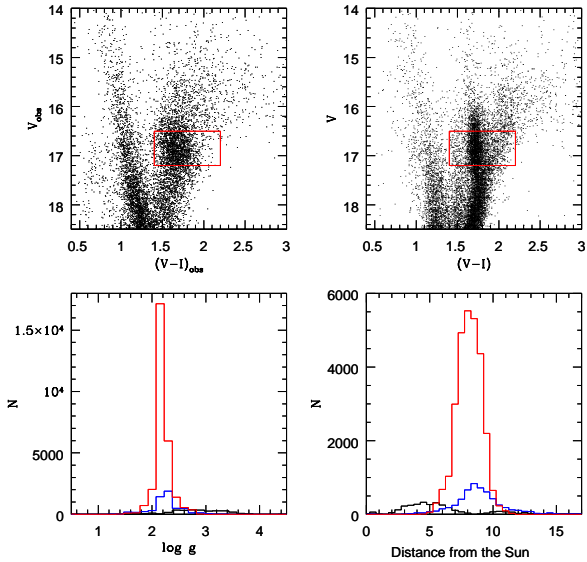}    
  \caption{The two upper panels show the (V,(V-I)) colour-magnitude
  diagrams of our field in Baade's window, as observed ({\em left})
  and simulated by the Besan\c con model ({\em right}). The red boxes
  in these panels show the sample selection. On the lower part of the
  figure, the distributions in gravities ({\em left}) and distances
  ({\em right}) of model stars falling in the selection box are
  presented, separated into their parent populations: bulge (red),
  thin disc (black) and thick disc (blue). The low level of expected
  contamination of our bulge red clump sample is evident on these
  distributions. Note also that, while the thin disc contaminants lie
  in the foreground, thick disc contaminants are located in the very inner
  regions of the Galaxy, at distances similar to the bulge red clump
  population.} 
\label{contamination_clump}
\end{figure}

We estimated the expected contamination of our bulge red clump sample
by foreground (and background) Galactic disc(s) stars and halo stars,
using the latest population synthesis Besan\c{c}on model
\citep{Robin2003,Picaud2004}. A $25\arcmin \times 25\arcmin$ field, centred on
Baade's Window, was simulated without including the reddening. The
latter was added independently of the Besan\c con model with the
following prescription based on best reproducing the locus of the CMD
in this region (position of the foreground disc main-sequence and red
clump sequences): for stars located beyond 2 kpc from the Sun
(including the bulge stars) we applied the same reddening as the one
used to compute the photometric temperature
(Sect.~\ref{Determine_stellar_param}, i.e. $\rm E(B-V)=0.55$), while for
stars within 2 kpc from the Sun (mostly thin disc), a reddening
proportional to the distance (to the Sun) was applied ($\rm
A_{V}=0.85/$kpc). Further shifts of the order of 0.1--0.2 mag had to be
applied to both V and V-I in order to well reproduce the position of
the bulge clump in the observed CMD. As shown in
Fig. \ref{contamination_clump}, the simulated CMD shows a global
morphology in agreement with the observed one although the colour distribution of the simulated CMD is narrower than the observed, owing to the combined effects of our ignoring the differential reddening within the field, and neglecting the photometric errors in the simulation.

In the sample selection box ($\rm 16.5 < V < 17.2$ et $\rm 1.4 < V-I <
2.2$), the total number of simulated stars is of $\sim 31200$ stars,
among which the vast majority are bulge stars (28383). The resulting
contamination predicted by the Besan\c{c}on model for our sample is of
9.3\%: 5.9\% from the thick disc, 3.3\% from the thin disc and 0.1\%
from the halo. For the sample of 219 bulge clump stars, this
corresponds to 7$\rm\pm$3 stars of the thin disc and 13$\rm\pm$4 stars
of thick disc. It can be appreciated from these low numbers that our
red clump sample should be very little contaminated by any other
intervening Galactic populations. 

As illustrated in
Fig.~\ref{contamination_clump}, the thick disc stars that enter our
selection box are mainly giants, with temperatures and gravities
similar to those of the red clump stars, and their large majority are
located beyond 7kpc from the Sun (i.e. in the inner parts of the
Galaxy). For these thick disc contaminating stars, the analysis
method developed for the red clump sample and in particular the
photometric gravity computed assuming that all the stars are located
at a distance of 8~kpc, can also be applied and will provide an
unbiased [Fe/H] estimation for these stars.  
The thin disc stars on the other hand, are also giants, but located in
majority  3-6~kpc from the Sun, and therefore have significantly higher
gravities than those of the red clump stars. This difference in \grav\
reaches 0.4 to 1~dex, which in turn leads to biases in the derived
[Fe/H] (with our method assuming that all stars are located in the
bulge, see Sec. \ref{Determine_stellar_param}) of approximately $-0.2$ to $-0.25$~dex. Whereas thick disc
contaminants from our sample could be readily identifiable from their
metallicity alone (assuming we know the metallicity distribution of
the thick disc in the inner regions of the galaxy), thin disc
contaminants are not expected to be easily identifiable from their
position in the metallicity distribution.  

The Besan\c{c}on model makes a number of hypothesis however, 
which may impact the estimates of contamination that we make 
here. One of them is the fact that the thin disc density is assumed to 
decrease sharply in its central 2\,kpc (it has a {\it hole} in the centre), 
which has been adjusted (together with its scalelength) to reproduce star 
counts towards  the bulge from the DENIS experiment, as explained in 
\citet{Picaud2004}.
This is the main reason for the little contamination by the thin disc 
predicted here: would the thin disc continue with an exponential profile 
all the way to the galactic centre, the RGB and RC stars in the central 
1-2kpc of the galaxy would contribute this contamination. On the other 
hand, observational evidence of this {\it hole} is quite compelling, 
from star counts, from the observation of a such a void in the gas 
surrounded by a molecular gas ring, and the compelling evidence 
for a stellar bar (\citep{Babusiaux2005,Nishiyama2006,Rattenbury2007a}) 
that would have swept much of the matter in these inner regions. 
More importantly perhaps, constraints on the density profile of the inner 
parts of the thick disc are very poor: the Besan\c{c}on model assumes 
a simple exponential profile, with no {\it hole} in the centre. This results in the major part of the predicted contamination to our sample to originate in the inner regions of the thick disc: this prediction is therefore rather poorly constrained.

Metallicity distributions assumed in the model also affect the position 
of the stars in each population in the colour magnitude diagram, and can 
shift stars in or out of our observed selection box. The thin disc is 
assumed to host a metallicity gradient of $\rm -0.07dex.kpc^{-1}$ (as 
constrained from the young stars and HII regions), 
which brings the innermost parts of the disk (around 2kpc from the centre, 
or 6kpc from the sun) to supersolar metallicities. If the inner thin disc 
had no hole, thin disc RGB and RC stars in the central 1-2kpc of the galaxy 
would have colours and magnitudes (and metallicities) very close to our 
sample's. However, the main-sequence counterpart of this younger population 
have so far eluded deep colour magnitude diagrams of low-reddening regions 
\citep{Clarkson2008}. The thick disc contribution is again the most uncertain 
part of the Besan\c{c}on model prediction, as there are currently no observational constraints regarding its radial metallicity gradient (or absence of). Would 
the thick disc in the model have a radial gradient, it may bring a few more 
stars in the selection box. However, even with a metallicity gradient, it is unlikely that thick disc stars will contaminate the high metallicity end of 
the sample, but would rather be found around metallicities of $-0.5$ or below.
  
\section{Determination of stellar parameters}\label{Determine_stellar_param}

\subsection{Photometric parameters}\label{photo_param}

Photometric temperatures were determined from the V-K, V-H and V-J colours using the calibrations of \cite{Ramirez2005} and their extinction law with $\rm E(B-V)=0.55$ to correct for reddening. 

The photometric gravities were calculated from the following equation:
   \begin{eqnarray}
\rm
log\left(\frac{g}{g_{\odot}}\right)=log\left(\frac{M}{M_{\odot}}\right)-0.4\left(M_{Bol,\odot}-M_{Bol,\ast}\right)+4log\left(\frac{T_{eff}}{T_{eff,\odot}}\right) 
\label{calcul_grav}
\end{eqnarray}

with $\rm M_{Bol,\odot}=4.72$ mag, $\rm T_{eff,\odot}=5770$ K, $\rm
logg_{\odot}=4.44$ dex for the Sun and $\rm M=0.8 M_{\odot}$ for the
bulge stars. The bolometric magnitude was computed from the V
magnitude and the bolometric correction calibrations of
\citet{Alonso1999} assuming that all stars were bulge members, at 8
kpc from the Sun \citep{Reid1993}. 

Despite the use of infrared bands, both photometric \Teff\ and \grav\
values are affected by the presence of the differential reddening but
show a different sensitivity. In the Baade's Window, the reddening
spread is of the order of 0.15 mag on V-I leading to
$\rm\sim$200K uncertainties on \Teff\ and $\rm\sim$0.05 dex on
\grav. Therefore the photometric temperature (mean value from the V-K,
V-H and V-J colours) was only used as an initial value in the stellar
parameter determination procedure (see Sect. \ref{FINAL_PARAM}). The
main source of uncertainty in the \grav\ determination is the distance
of the star. Assuming that all sample stars are at the mean bulge
distance to the Sun can lead to an error up to 0.25 dex on
\grav. However, on the whole sample, the previous uncertainty value
remain lower than those expected from a \grav\ deduced from the
ionisation equilibrium. This is mainly due to uncertainties on the
\ion{Fe}{II} abundances coming from uncertainties in the equivalent widths
measurements (continuum placement, blends) and on the \GF\ values of
the \ion{Fe}{II} lines (see \cite{Lecureur2007}). At GIRAFFE
resolution, the previous source of uncertainties become larger and the
photometric gravity was used as a final value. We also note that the
\Teff\ value influences the photometric \grav: a change of
$\rm\sim200$ K can lead to changes on \grav\ up to 0.15 dex for the
cooler stars. Then, the \grav\ value were recomputed at each step of
the stellar parameters determination procedure taking into account the
change on \Teff.  

\subsection{Spectroscopic parameters}

\subsubsection{EWs measurement, atmospheric models and codes}

The equivalent widths (EWs) for selected \ion{Fe}{I} and \ion{Mg}{I}
lines were measured using the automatic code DAOSPEC\footnote{DAOSPEC
  has been written by P.B. Stetson for the Dominion Astrophysical
  Observatory of the Herzberg Institute of Astrophysics, National
  Research Council, Canada.} \citep{Stetson2008}. The
synthetic spectra were computed using the LTE spectral analysis code
``turbospectrum'' (described in \citealt{plezcode}) and the abundances
from EWs were derived using the Spite programs \citep[][ and
  subsequent improvements over the years]{Spite1967} with for both
codes the new MARCS models \citep{OSMARCS}.   

\subsubsection{The iron linelist}

The iron linelist from \citet{Lecureur2007} was used for the stellar
parameter determination. This linelist was established in the
following way: as a starting point, we used the linelist from
\citet{Zoccali2004} with \GF\ values extracted from the NIST
database\footnote{Available: http://physics.nist.gov/asd3}
\citep{NIST}. By computing synthetic spectra in the stellar parameters
range of the red clump stars, we rejected the iron lines blended up to
10\% of the equivalent width. Finally, for the remaining iron lines,
the \GF\ values were adjusted in order that each line gives an
abundance of 0.30 dex from the EW measured on the observed spectrum of
\MuLeo\ with the following stellar parameters:  \Teff~$=4540$~K,
\grav~$=2.3$~dex, and \VT~$=1.3$ km\,s$^{-1}$. The spectrum of \MuLeo\
was obtained at the Canada-France-Hawaii Telescope with the ESPaDOnS
spectropolarimeter (resolution R$\rm\sim$80000 and S/N per pixel
$\rm\sim500$) and processed using the ``Libre ESpRIT" data reduction
package \citep{Donati1997}. 

Restricted to the wavelength ranges of H13+H14, the linelist contains
48 \ion{Fe}{I} lines (92 initially). We checked the consistency of
this reduced linelist on Arcturus and the Sun but also on the 12 red
clump stars observed with GIRAFFE and UVES. The same parameters were
found for Arcturus with the reduced and total linelists :
\Teff~$=4300$~K, \grav~$=2.5$~dex, \VT~$=1.5$ $\rm km s^{-1}$ and [Fe/H]~$=-0.52$
dex. For the Sun, we obtained \grav~$=4.4$~dex, \VT~$=1.5$
km\,s$^{-1}$ but \Teff$=5877$~K, ie increased by 100~K to fullfill the
excitation equilibrium criterium. However, this has no impact on the
deduced metallicity. For the 12 red clump stars, a new determination
of the stellar parameters has been performed with the reduced linelist
from the EWs measured on the UVES spectra with the same method as
described in \cite{Lecureur2007}. On the mean, the difference on the
\Teff\ values are null with a dispersion of 100 K, lower than the
uncertainties. The \VT\ resulting from the use of the reduced linelist
are slightly lower (0.1 km\,s$^{-1}$) than the ones found with the
total linelist which translates into [Fe/H] values higher (0.1 dex)
than the previous ones. However, these systematics are smaller than
the uncertainties found on the individual values and let us conclude
that the stellar parameters determination from the reduced line is
reliable.      

\subsubsection{The microturbulence velocity determination}

In \cite{Lecureur2007}, \VT\ was determined so that lines of different
observed EWs give the same iron abundance. However, the observed EW
values are sensitive to uncertainties due to either the quality of the
spectra (S/N, resolving power) or the analysis method (continuum
placement, FWHM fittings, ... ) which translates to correlated
uncertainties on the corresponding [\ion{Fe}{i}/H]. This correlation between
[\ion{Fe}{i}/H] and observed EWs uncertainties leads to a bias towards higher
\VT\ as mentioned by \citet{Magain1984}.  

\begin{figure}
   \centering
   \resizebox{\hsize}{!}{\includegraphics [angle=-90]{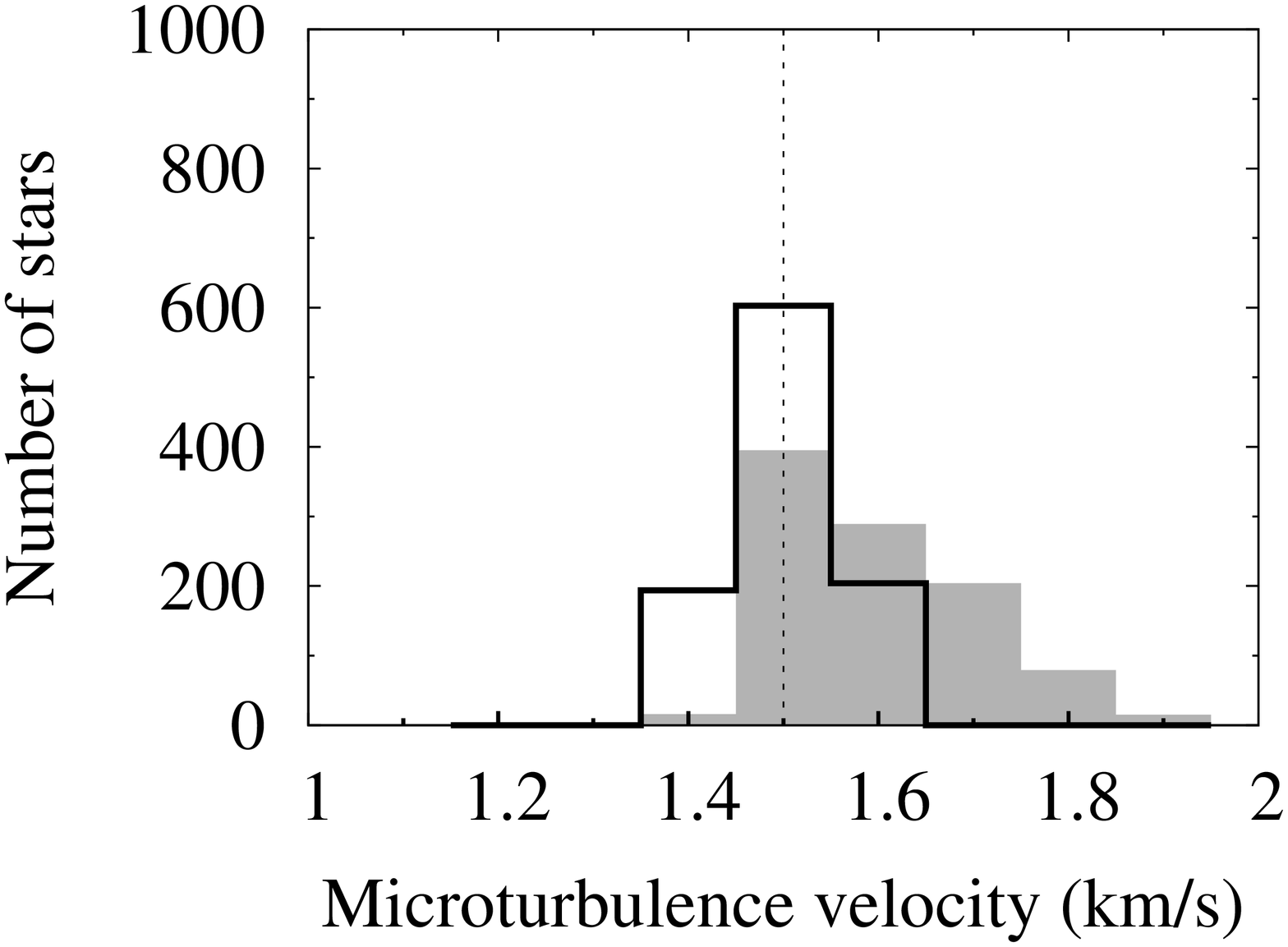}
   \includegraphics [angle=-90]{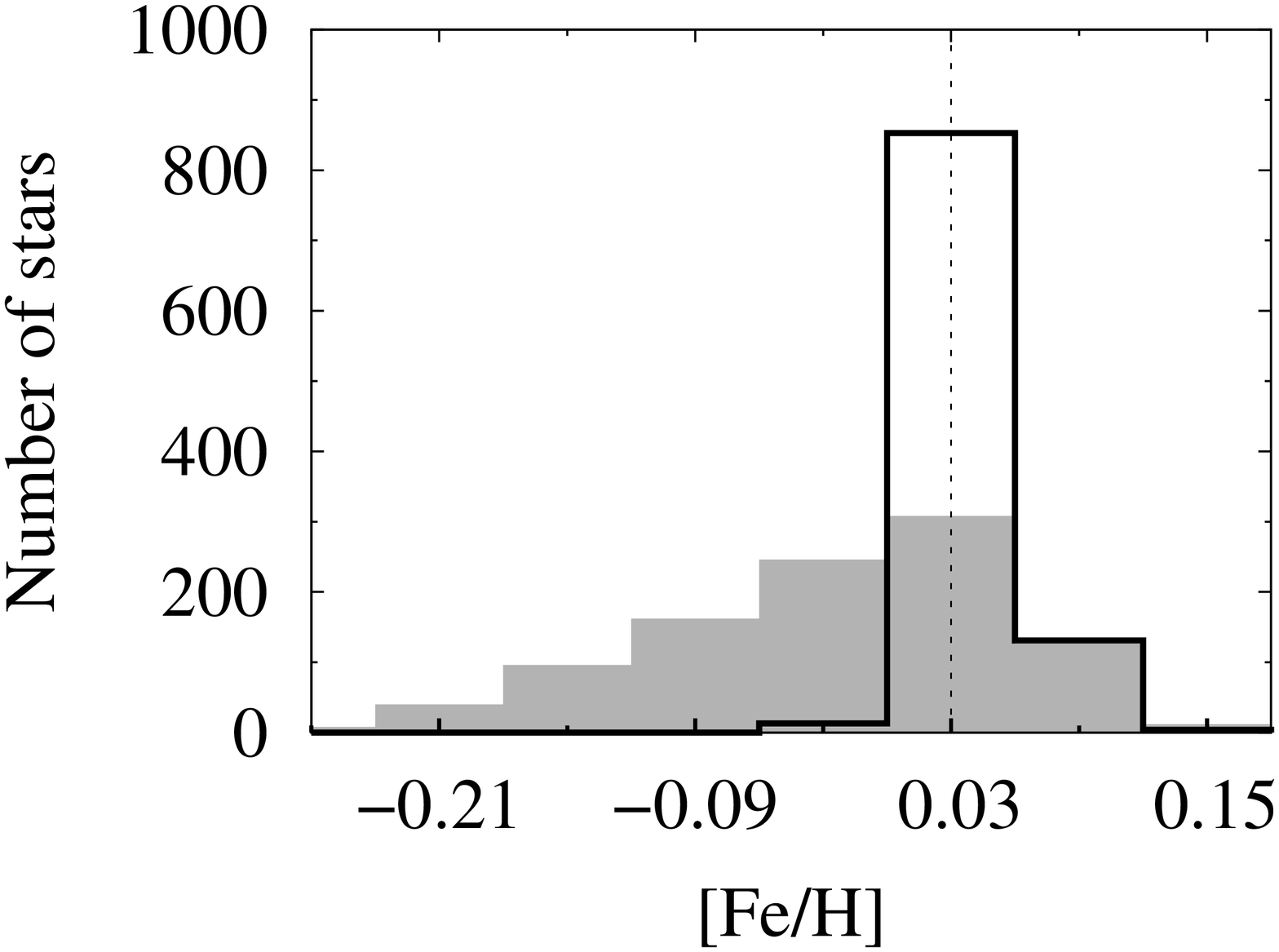}}
    \caption{Results of the simulations for the star BWc-2 (\Teff\
    $\rm =4656$ K, \grav\ $\rm =2.30$ dex, \VT\ $\rm =1.5$
    km\,s$^{-1}$ and $\rm [\ion{Fe}{I}/H]=0.03$). \textit{Left panel:}
    Histogram of \VT\ values found with the observed EWs (shaded histogram)
    and theoretical EWs (black line). \textit{Right panel:} Histogram of
    the mean $\rm [\ion{Fe}{I}/H]$ abundances. The correlation between
    uncertainties in observed EWs and the \ion{Fe}{I} abundances leads
    to a systematic overestimate of \VT\ determined from the
    abundances versus observed EWs plot. This overestimate is not
    significant anymore with the use of theoretical EWs.} 
         \label{graphSimu}
   \end{figure}

In order to estimate this effect on the GIRAFFE sample, we performed
simulations of the observed EWs and measured \VT\ for each
simulation. This procedure was applied to the star BWc-2 (a star
representing well our sample: \Teff~$\rm =4656$ K, \grav\ $\rm =2.30$
dex, \VT\ $\rm =1.5$ km\,s$^{-1}$ and $\rm [\ion{Fe}{I}/H]=0.03$).  We
simulated 1000 stars by randomly drawing a set of \ion{Fe}{I} lines
EWs ($\rm EWs_{rand}$) around the expected EWs ($\rm EWs_{exp}$) with
a standard deviation $\rm\sigma=\delta_{EWs}$ (given by DAOSPEC). For
each 1000 simulated stars, \ion{Fe}{I} abundances
($\rm[\ion{Fe}{i}/H]_{rand}$) and corresponding uncertainties were computed
from the random EWs using the stellar atmosphere model adopted for
BWc-2. And \VT\ was modified to minimise the slope p of
$\rm[\ion{Fe}{i}/H]_{rand}$ versus $\rm EWs_{rand}$ such that $\rm
\left|p\right|<\sigma_{p}$, with $\rm \sigma_{p}$ the uncertainty on
p. The simulation results are shown in Fig. \ref{graphSimu}. The
systematic effect on the determination of \VT\ using the observed EWs
is clearly illustrated by the \VT\ distribution histogram. In the
mean, \VT\ has to be increased by $\rm 0.10\pm0.11$ km\,s$^{-1}$
leading to a corresponding systematic effect on the mean $\rm
[\ion{Fe}{I}/H]$ abundance which decreases by $\rm 0.06\pm0.08$ dex.  


 


This systematic effect can be suppressed using an abscissa linked to
the observed EW but not affected by random errors. \citet{Magain1984}
suggested to use the expected EWs (computed from the adopted stellar
atmophere model and some assumed values of \VT) and proposed a scheme
where the slope of $\rm[\ion{Fe}{i}/H]$ versus $\rm EWs_{exp}$ is minimised to
deduce \VT. Such \VT\ values are then free from bias EW abundance
error correlations. This scheme was applied to the previously
simulated sample. As clearly illustrated by the Fig. \ref{graphSimu},
the systematic effect on \VT\ is not significant anymore:  the mean
\VT\ value is of $\rm 1.50\pm0.06$ km\,s$^{-1}$ and the corresponding
mean $\rm [\ion{Fe}{I}/H]$ value is of $\rm 0.02\pm0.03$ dex. 

However, this method can introduce another type of bias on \VT\ (and
thus on [Fe/H]) if the $\rm EWs_{exp}$ are computed from an atmosphere
model that is far from reality. In particular, [Fe/H] needs to be
known to $\rm\sim0.07$ dex. For this reason, [Fe/H] was estimated as
well as possible before \VT\ was finally constrained.  

\subsection{Final stellar parameters determination}\label{FINAL_PARAM}

Starting with the photometric temperature, the final \Teff\ was
determined iteratively by requiring the excitation equilibrium on
\ion{Fe}{I} lines (no variation of [\ion{Fe}{i}/H] with the excitation
potential of the line $\rm \chi_{ex}$. \VT\ was constrained requiring no trend in
the \ion{Fe}{I} abundance as a function of the expected EW of the lines. At each step of
the iteration on \VT, we checked that the metallicity of the atmophere
model was the same as the one derived from the average \ion{Fe}{I}
abundance. As explained in Sect. \ref{photo_param}, the photometric
gravity was adopted as a final value and recomputed at each step of
the iteration taking into account the change in \Teff\ and
[Fe/H]. Stellar parameters and [Fe/H] values were finally obtained for
219 red clump stars of the initial sample (228 stars), and are
reported in Table \ref{Table_ParamAbund}, available at the CDS
(the first lines of the table are reproduced in the
printed version). 

\begin{figure}
   \centering
   \resizebox{\hsize}{!}{\includegraphics [angle=-90]{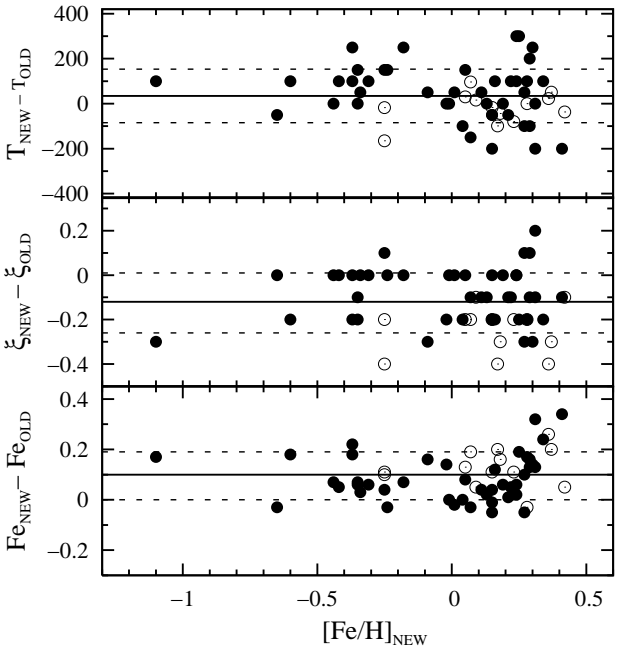}}
    \caption{Comparison of the stellar parameters ([Fe/H], \VT\ and
    \Teff) obtained with the new \VT\ determination method (NEW) and
    those published in \citet{Lecureur2007} (OLD). The black dots
    represent the RGB stars and the open circle the red clump
    stars. The black line shows the mean difference and the dotted
    lines the $\rm\pm1\sigma$ dispersion.  With the new method,
    the \VT\ values are systematically lower which implies [Fe/H]
    systematically higher.} 
         \label{diff_param_old_new_uves}
   \end{figure}

This method was also applied to the UVES sample of
\citet{Lecureur2007}, using their full linelist,
 leading to new stellar parameters for each star
of the sample. Fig. \ref{diff_param_old_new_uves} shows the comparison
between the new stellar parameters obtained with the new \VT\
determination method and the ``old" ones. There is no significant
change in \Teff\ values, the difference between old and new values is
of $\rm-34\pm119$ K. As expected with this new method, \VT\ values are
on the mean found systematically lower of $\rm0.12\pm0.14$
km\,s$^{-1}$ which translates into [Fe/H] values systematically higher
of $\rm0.10\pm0.09$ dex. Compared to the whole sample, the red clump
stars are more sensitive to the change of \VT\ determination with a
mean difference between old and new values which reaches
0.25~km\,s$^{-1}$. This can be explained by lower S/N for these stars
compared to the total sample and consequently higher uncertainties on
the EWs and corresponding \ion{Fe}{I} abundances making these stars
more sensitive to the bias affecting the \VT\ determination from the
observed EWs.  

\begin{table*}[!htbp]
\begin{center}
\caption{Adopted stellar parameters and measured [Fe/H] and [Mg/H] in
  the sample stars.}
\label{Table_ParamAbund}
\begin{tabular}{llllrrrrrr}
\hline
\hline
\multicolumn{1}{c}{ID} & 
\multicolumn{1}{c}{\Teff} & \multicolumn{1}{c}{\grav}  & \multicolumn{1}{c}{\VT}  & 
\multicolumn{1}{c}{[Fe/H]}  & \multicolumn{1}{c}{$\sigma_{\rm Fe}$(low)} & \multicolumn{1}{c}{$\sigma_{\rm Fe}$(up)}  & 
\multicolumn{1}{c}{[Mg/H]}  & \multicolumn{1}{c}{$\sigma_{\rm Mg}$(low)} & \multicolumn{1}{c}{$\sigma_{\rm Mg}$(up)}  \\
& K & dex & $\rm km s^{-1}$ & dex & dex & dex & dex & dex & dex \\
\hline 
BWc-1  & 4528 & 2.09 & 1.4 &  0.29 & 0.14 & 0.13 &  0.34 & 0.08 & 0.08 \\
BWc-2  & 4556 & 2.25 & 1.4 &  0.04 & 0.10 & 0.21 &  0.29 & 0.10 & 0.10 \\
BWc-3  & 4713 & 2.23 & 1.5 &  0.33 & 0.17 & 0.28 &  0.42 & 0.15 & 0.15 \\
BWc-4  & 4912 & 2.28 & 1.4 &  0.01 & 0.10 & 0.19 &  0.26 & 0.10 & 0.10 \\
BWc-5  & 4636 & 2.21 & 1.5 &  0.58 & 0.12 & 0.33 &  0.63 & 0.14 & 0.14 \\
BWc-6  & 4769 & 2.19 & 1.2 & -0.17 & 0.17 & 0.30 & -0.00 & 0.19 & 0.19 \\
BWc-7  & 4769 & 2.19 & 1.2 & -0.17 & 0.17 & 0.30 & -0.00 & 0.19 & 0.19 \\
BWc-8  & 4810 & 2.31 & 1.4 &  0.49 & 0.15 & 0.17 &  0.47 & 0.09 & 0.09 \\
BWc-9  & 4685 & 2.21 & 1.5 &  0.17 & 0.12 & 0.24 &  0.21 & 0.12 & 0.12 \\
BWc-10 & 4690 & 2.13 & 1.5 & -0.13 & 0.14 & 0.24 &  0.20 & 0.11 & 0.11 \\
...    &  ... & ...  & ... & ...   & ...  & ...  & ...   & ...  & ...\\
\hline
\end{tabular}
\end{center}
\end{table*}

\subsection{The stellar parameters uncertainties}\label{stellar_param_error}

For each star of the red clump sample, an estimate of the uncertainty
on [Fe/H] has been computed from the uncertainties on the stellar
parameters. The uncertainties on \Teff\ and \VT\ have been evaluated
from the uncertainties on the \ion{Fe}{I} line EWs given by DAOSPEC
which translate into uncertainties in \ion{Fe}{I} abundances derived
from the EWs. These individual uncertainties on the \ion{Fe}{I} lines
were considered in the least squares line fit to the points in the
($\rm \chi_{ex},[\ion{Fe}{I}/H]$) plane to fix \Teff\ and in the ($\rm
EWs,[\ion{Fe}{I}/H]$) plane to fix \VT\ and are contained in the error
on the line slope.  

Keeping a fixed \VT, \Teff\ upper and lower allowed boundaries were
estimated by letting the slope p reaches its 1 $\rm\sigma$
uncertainty:  $\rm p\pm\sigma_p$. The thus determined boundaries are
called \Teff$\rm_{,up}$ and \Teff$\rm_{,lo}$ and the corresponding
uncertainties on \Teff\ $\rm \sigma_{up}$(\Teff) and ($\rm
\sigma_{lo}$(\Teff) respectively. Note that these temperature
uncertainties need not to be symmetric. In fact, for the 219 red clump
stars, we found:  $\rm \sigma_{sup}$(\Teff)$\rm=180\pm90$ K and $\rm
\sigma_{inf}$(\Teff)$\rm=220\pm90$ K and half of the sample shows
asymmetric uncertainties on \Teff\ which can reach 200 K. Lower ($\rm
\sigma_{lo}$(\VT)) and upper ($\rm \sigma_{up}$(\VT)) uncertainty
values on \VT\ have been estimated with the same principle fixing the
\Teff\ value. Contrary to the \Teff's, these uncertainties are quasi
symmetric with $\rm |\sigma_{inf}$(\VT)$\rm |=|\sigma_{sup}$(\VT)$\rm
|=0.21\pm0.08$ km\,s$^{-1}$.

\begin{figure}
   \centering
   \resizebox{\hsize}{!}{\includegraphics [angle=-90]{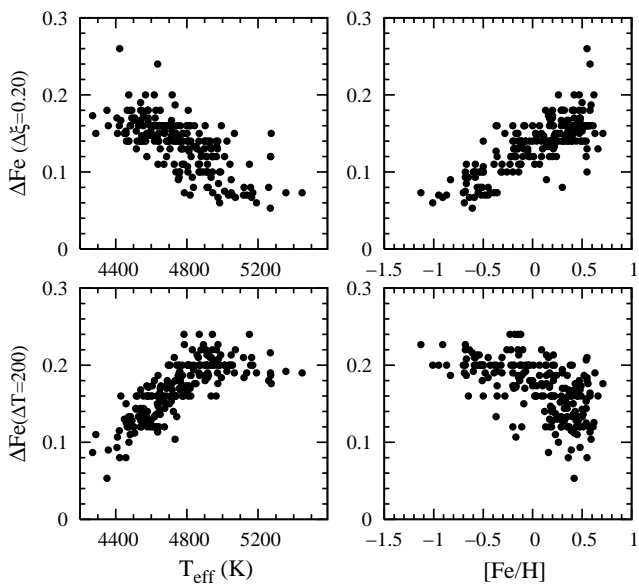}}
    \caption{Variation in the resulting [Fe/H] with increasing \Teff\
    by 200~K (\textit{lower panels}) and \VT\ by 0.2~km$\rm s^{-1}$
    (\textit{upper panels}) as a function of the initial \Teff\
    \textit{(left panels}) and of the initial [Fe/H] value
    (\textit{right panels}). The hotter and less metallic stars are
    the most sensitive to a modification of \Teff\ whereas the cooler
    and more metallic stars are the most sensitive to a modification
    of \VT.} 
         \label{diff_fe_due_to_diff_teff}
   \end{figure}

A modification in \Teff\ (or in \VT) value induces a change in [Fe/H]
which varies from star to star depending on the initial stellar
parameters but also on the quality of the EW measurement (S/N ratio,
continuum placement, ... ). As illustrated by
Fig. \ref{diff_fe_due_to_diff_teff}, changing \Teff\ by 200 K implies
a change of [Fe/H] around 0.10 dex for the coolest and more metallic
stars of the sample and around 0.20 dex for the hotter and less
metallic stars. Changes of \VT\ by 0.2 km$\rm s^{-1}$ lead to similar
variations in the [Fe/H] value: around 0.10 dex for the coolest and
more metallic stars of the sample and around 0.16 dex for the hotter
and less metallic stars. Overall in our 219 clump stars sample, the
uncertainties on [Fe/H] associated to \Teff\ uncertainties are
of $\rm 0.15\pm0.07$ dex on the mean, and those associated to
\VT\ uncertainties, of $\rm 0.16\pm0.06$ dex. The individual values
are reported in the Table \ref{Table_errors}, available at the CDS
in its entirety (the first lines of the table are reproduced in the
printed version).    

The uncertainty associated with [Fe/H] were computed with the
formalism described in \citet{McWilliam1995} taking into account the
errors on the model atmosphere stellar parameters \Teff\ and \VT\ and
the errors on the \ion{Fe}{I} line EW measurements. Errors on [Fe/H]
coming from the \grav\ value or the metallicity adopted to compute the
stellar atmosphere model were found to be negligible compared to the
other sources. Finally, the variance on [Fe/H], $\rm \sigma^2(Fe)$,
can be written as: 

\begin{eqnarray}
\rm \sigma^2(Fe) &=& \left(\frac{\partial\epsilon}{\partial W}\right)^2\sigma^2_{W}
                   +\left(\frac{\partial\epsilon}{\partial T}\right)^2\sigma^2_{T}
                   +\left(\frac{\partial\epsilon}{\partial \xi}\right)^2\sigma^2_{\xi}
                     \nonumber\\
                 && +2\left(\frac{\partial\epsilon}{\partial T}\right)
                   \left(\frac{\partial\epsilon}{\partial \xi}\right)\sigma_{T,\xi}, 
\label{calcul_error_fer}
   \end{eqnarray}

with $\epsilon$ the Fe abundance, 
$\rm \sigma^2_{W}$, $\rm \sigma^2_{T}$ and $\rm \sigma^2_{\xi}$
the variance in the EWs measurement, \Teff\ and \VT, respectively and
$\rm \sigma_{T,\xi}$, the covariance between \Teff\ and \VT. The
covariances between the EW measurements and the stellar parameters
\VT\ or \Teff\ are null because these variables are independent. The
terms $\rm \left(\frac{\partial\epsilon}{\partial
  T}\right)^2\sigma^2_{T}$ and $\rm
\left(\frac{\partial\epsilon}{\partial \xi}\right)^2\sigma^2_{\xi}$
were computed as explained in the previous paragraph. The variance in
the EW measurements and the associated uncertainty on [Fe/H] were
estimated from the line to line dispersion divided by the number of
lines. The covariance between \Teff\ and \VT\ were approximated by
measuring the variation in the slope fixing \Teff\ (resp. \VT) due to
changes in \VT\ (resp. \Teff).  
 
\begin{figure}
   \centering
   \resizebox{\hsize}{!}{\includegraphics [angle=-90]{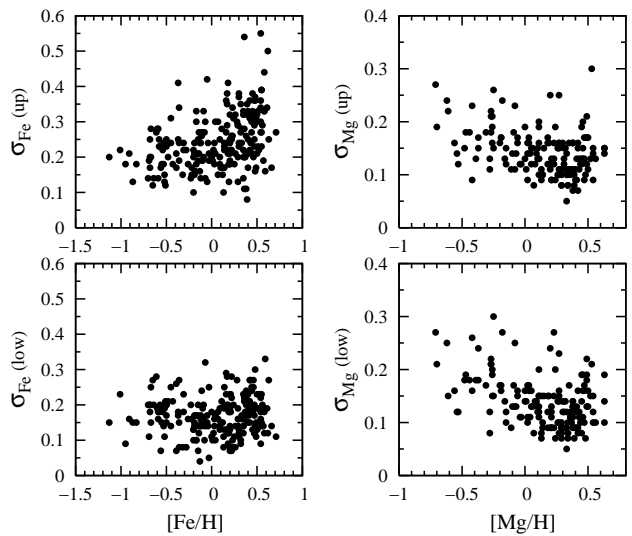}}
    \caption{\textit{Left panels:} upper and lower errors on [Fe/H]
    as a function of [Fe/H] for the 219 red clump stars. The upper
    values increase with the metallicity with mean values around 0.3
    dex for the most metallic stars. The lower values show no trend
    with [Fe/H] with a median value of 0.17 dex for the whole
    sample. \textit{Right panels:} upper and lower errors on [Mg/H] as
    a function of [Mg/H] for the 162 red clump stars selected from the
    total sample. At low [Mg/H], the abundance errors are dominated by the line measurement rather than stellar parameter uncertainties because} the Mg triplet lines become very weak.
         \label{errors_fe}
   \end{figure}

The total upper and lower uncertainty values on [Fe/H] and [Mg/H] computed from
equation~\ref{calcul_error_fer} are reported in Table~\ref{Table_ParamAbund}, 
while Table~\ref{Table_errors} details the different contributions of the stellar 
parameter uncertainties ($\sigma_{\xi}$ and $\sigma_{\rm T}$) propagated onto the 
abundances ($\sigma_{\xi}(\rm Fe)$). On the
whole red clump sample (219 stars), and as illustrated in Fig.~\ref{errors_fe},
 the lower values range from 0.04
to 0.35 dex with a median value of $\rm0.17\pm0.05$ and show no trend
with metallicity. The upper values are systematically higher with a
median value of $\rm0.24\pm0.08$ and show an increase with the
metallicity reaching 0.5 dex for supersolar metallicity stars. 

\begin{table*}[!htbp]
\begin{center}
\caption{Errors in stellar parameters and associated errors on Fe for the sample stars.}
\label{Table_errors}
\begin{tabular}{llllcccccccccc}
\hline
\hline
&&&&& \multicolumn{4}{c}{lower} && \multicolumn{4}{c}{upper}\\
\cline{6-9} \cline{11-14}
\multicolumn{1}{c}{ID} & 
\multicolumn{1}{c}{\Teff} & \multicolumn{1}{c}{\grav}  &
\multicolumn{1}{c}{\VT}  &  \multicolumn{1}{c}{[Fe/H]} &
\multicolumn{1}{c}{$\sigma_{\xi}$}  & \multicolumn{1}{c}{$\sigma_{\xi}(\rm Fe)$} &
\multicolumn{1}{c}{$\sigma_{\rm T}$}  & \multicolumn{1}{c}{$\sigma_{\rm T}(\rm Fe)$} &&
\multicolumn{1}{c}{$\sigma_{\xi}$}  & \multicolumn{1}{c}{$\sigma_{\xi}(\rm Fe)$}&
\multicolumn{1}{c}{$\sigma_{\rm T}$}  & \multicolumn{1}{c}{$\sigma_{\rm T}(\rm Fe)$} \\
& \multicolumn{1}{c}{K} & \multicolumn{1}{c}{dex} &
\multicolumn{1}{c}{$\rm km s^{-1}$} & \multicolumn{1}{c}{dex} & 
\multicolumn{1}{c}{$\rm km s^{-1}$} & \multicolumn{1}{c}{dex} &
\multicolumn{1}{c}{K} & \multicolumn{1}{c}{dex} &&
\multicolumn{1}{c}{$\rm km s^{-1}$} & \multicolumn{1}{c}{dex} &
\multicolumn{1}{c}{K} & \multicolumn{1}{c}{dex} \\
\hline 
BWc-1  & 4528 & 2.09 & 1.4 & +0.29 & -0.10 & +0.07 & +150 & +0.10 && +0.20 & -0.16 & -200 & -0.10 \\
BWc-2  & 4556 & 2.25 & 1.4 & +0.04 & -0.20 & +0.19 & +100 & +0.08 && +0.10 & -0.07 & -150 & -0.07 \\
BWc-3  & 4713 & 2.23 & 1.5 & +0.33 & -0.20 & +0.16 & +300 & +0.23 && +0.30 & -0.22 &  -50 & -0.03 \\
BWc-4  & 4912 & 2.28 & 1.4 & +0.01 & -0.20 & +0.19 & +100 & +0.10 && +0.10 & -0.06 & -100 & -0.09 \\
BWc-5  & 4636 & 2.21 & 1.5 & +0.58 & -0.30 & +0.30 & +300 & +0.17 && +0.10 & -0.12 & -400 & -0.16 \\
BWc-6  & 4769 & 2.19 & 1.2 & -0.17 & -0.20 & +0.15 & +300 & +0.26 && +0.20 & -0.13 & -200 & -0.15 \\
BWc-7  & 4769 & 2.19 & 1.2 & -0.17 & -0.20 & +0.15 & +300 & +0.26 && +0.20 & -0.13 & -200 & -0.15 \\
BWc-8  & 4810 & 2.31 & 1.4 & +0.49 & -0.10 & +0.10 & +150 & +0.13 && +0.10 & -0.07 & -200 & -0.14 \\
BWc-9  & 4685 & 2.21 & 1.5 & +0.17 & -0.20 & +0.20 & +150 & +0.13 && +0.20 & -0.14 & -200 & -0.15 \\
BWc-10 & 4690 & 2.13 & 1.5 & -0.13 & -0.20 & +0.15 & +200 & +0.18 && +0.30 & -0.21 & -200 & -0.14 \\
...    &  ... & ...  & ... & ...   & ...   & ...   & ...  & ...   && ...   & ...   & ...  & ...   \\
\hline
\end{tabular}
\end{center}
\end{table*}

We also mention that a mean error in [Fe/H] was also estimated by
Zoccali et al. (2008) by comparing the measured metallicities in the
50 stars (RGB and red clump stars of the \citet{Lecureur2007}'s
sample) observed both with GIRAFFE and UVES. They found no systematic
offset on the [Fe/H] values with a scatter of 0.16 dex, a value in
agreement with the previous estimates from the total GIRAFFE red clump
sample.

\subsection{Magnesium abundances}\label{method_measure_Mg}

The Mg abundances were determined from the 6319\AA\ triplet using
spectral synthesis fitting. The synthetic spectra were computed using
the LTE spectral analysis code ``turbospectrum" \citep[described
  in][]{plezcode} and the molecular and atomic linelists described in
\citet{Lecureur2007}. In short, we recall that the atomic linelist is
the one of the VALD database \citep[][ and references
  therein]{Kupka1999} with (i) solar astrophysical \GF\ -values for
the three lines of the Mg triplet and (ii) the $\rm \lambda$
6318.1~\AA\ \ion{Ca}{I} autoionisation line broadening adjusted in order to
reproduce the \ion{Ca}{I} dip in the Sun, Arcturus and \MuLeo. The
molecular line lists included in our syntheses are the following:
C$_2$ ($^{12}$C$^{12}$C and $^{12}$C$^{13}$C) Swan system (A-X); 
CN ($^{12}$C$^{14}$N and $^{13}$C$^{14}$N) red system (A-X); 
TiO $\gamma$ and $\gamma'$ systems \citet{Plez1998}.  In and near
the triplet, the \GF\-values of some CN lines were modified in order
to obtain the same Mg abundance from the three lines of the triplet in
the three reference stars. 

\subsubsection{In the UVES stars: new determination}\label{new_determination_uves}

Adopting the new stellar parameters and metallicities (see
Sect.~\ref{FINAL_PARAM}), the Mg abundances were determined for the
UVES stars as described in \citet{Lecureur2007} with a Ca abundance
computed from the following relation: $\rm
[Ca/Fe]=-0.20[Fe/H]+0.06$. This linear relation was obtained on the
UVES sample by a least square fit to the [Ca/Fe] values deduced from
EWs.  

   \begin{figure}[ht!]
   \centering
   \resizebox{\hsize}{!}{\includegraphics [angle=-90]{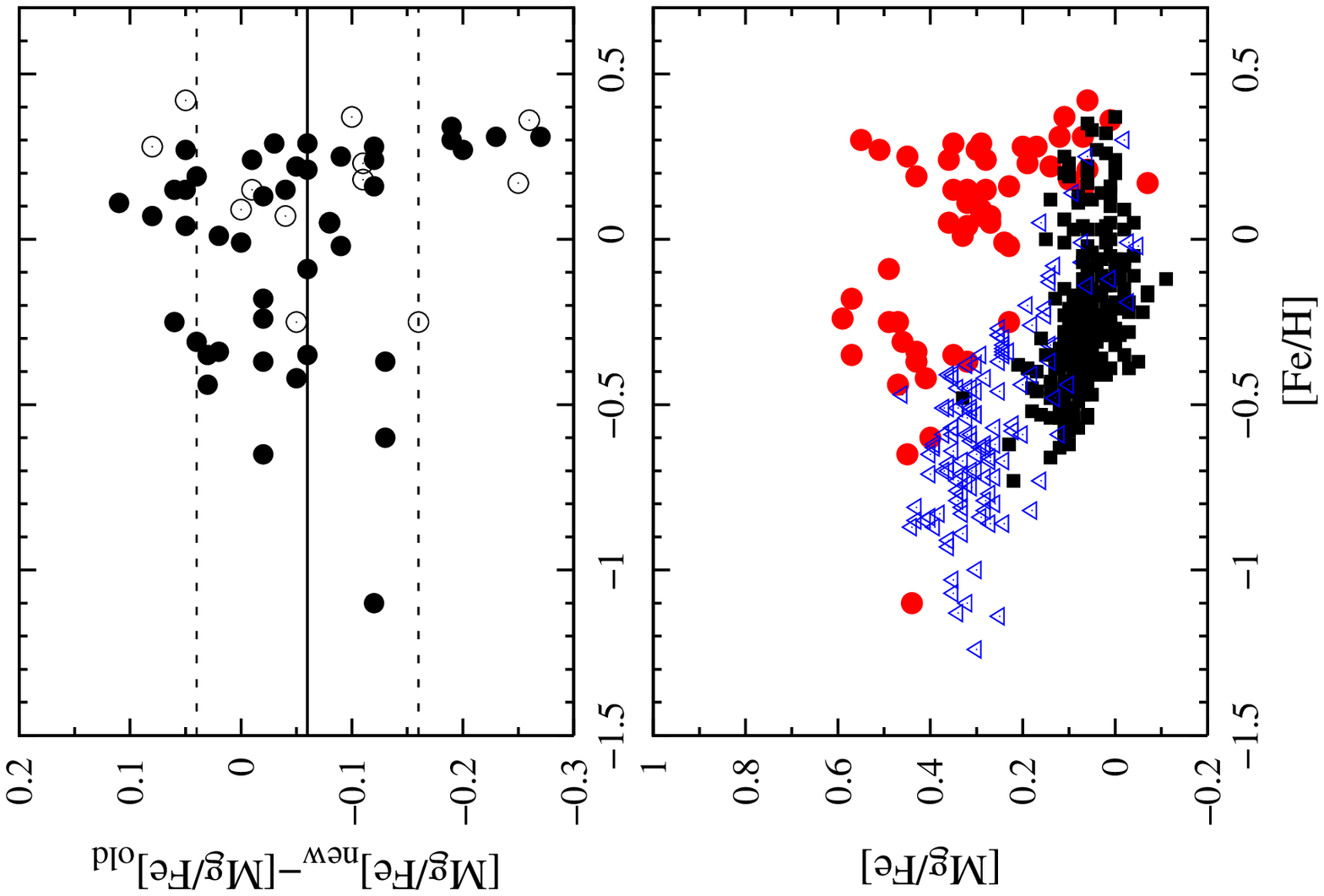}}
    \caption{\textit{Upper Panel}: Difference on [Mg/Fe] found
    for the UVES red clump (open circle) and RGB stars (full circle)
    with the old and new set of parameters as a function of the new
    metallicities. The black line shows the mean value of this
    difference (-0.06) and the two dotted lines shows the
    $\rm\pm1\sigma$ dispersion ($\rm\sigma=0.10$). 
    \textit{Lower Panel}: [Mg/Fe] in the UVES bulge stars
    (red points, computed with the new stellar parameters) compared
    with the [Mg/Fe] in the thin (black squares) and thick (blue
    triangles) discs stars from
    \cite{Bensby2004,Bensby2005,Reddy2006}. For metallicities from
    -0.4 to 0.1, the bulge stars show higher [Mg/Fe] values than in
    the two discs. } 
         \label{mg_uves}
   \end{figure}
 
The mean difference in [Mg/H] between the two determinations is very small
(as expected because [Mg/H] is not sensitive to a change of \VT):
$\rm[Mg/H]_{new}-[Mg/H]_{old}=0.03\pm0.06$ dex but becomes more
significant on [Mg/Fe] with
$\rm[Mg/Fe]_{new}-[Mg/Fe]_{old}=-0.06\pm0.10$ dex (see
Fig.\ref{mg_uves}). However, the global trend of [Mg/Fe] vs [Fe/H] is
almost not affected by this difference as illustrated by
Fig.\ref{mg_uves} which shows the new [Mg/Fe] in the bulge stars and
the [Mg/Fe] found in the thin and thick discs stars by
\cite{Bensby2004,Bensby2005} and \cite{Reddy2006}. In the lower
metallicities range ([Fe/H]$\rm<-0.4$), the bulge stars are not as
separated from the thick disc stars than they were in \cite[][see
  Fig.6]{Lecureur2007} with [Mg/Fe] values in bulge stars similar to
the highest values found in the thick disc stars. For
$\rm-0.4\lesssim[Fe/H]\lesssim0.1$, the bulge still shows higher
[Mg/Fe] values than those  found in both discs. At higher
metallicities, the dispersion has slightly decreased (this comes
mainly from our use of the linear relation to compute the Ca abundance
rather than relying on individual --more uncertain-- values for each
star) but remains high, with [Mg/Fe] values from thin disc values
($\rm[Mg/Fe]\sim0.1$) to 0.55. The conclusions drawn by
\cite{Lecureur2007} from the [Mg/Fe] values are still valid with the
     [Mg/Fe] computed from the new stellar parameters. The abundances
     of O, Al and Na in the UVES stars will also be updated using
     these new stellar parameters and be published and discussed in a
     future paper.  

\subsubsection{In the GIRAFFE stars}

From the total red clump sample (219 stars), we selected stars
according to their uncertainties on the [Fe/H] values in order to
exclude stars for which Mg would be measured with too large
uncertainties (mainly stars with low S/N spectra). Only stars with
$\rm\sigma_{up}(Fe)<0.30$ were kept providing a new subsample of 162
stars. To compute the synthetic spectrum, the abundances of C,N, O and
Ca are needed for each star of the GIRAFFE sample. As a result from
the decrease in spectral resolution, the C, N and O indicators
measurable in the UVES spectra are either nondetectable in the GIRAFFE
spectra or would lead to large uncertainties in the resulting
abundances, mainly due to uncertainties in the continuum placement
($\rm \sim0.5$ dex on [C/H] and $\rm \sim0.3$ dex on
[O/H]). Therefore, the synthetic spectra were computed with the mean
values, [C/Fe]=-0.04 and [N/Fe]=+0.43, found in the UVES stars
sample. The O abundance was deduced from the following linear
relation: [O/Fe]=-0.56[Fe/H]+0.22 fitted by least squares to the
[O/Fe] values found for the UVES stars sample of
\cite{Zoccali2006}. The Ca abundance was computed as in
Sect.~\ref{new_determination_uves}. 

Each observed spectrum was first normalised using the continuum found
by DAOSPEC. The continuum was then adjusted by a visual inspection of
a 10 \AA\ wavelength interval centred on the triplet. In order to
well identify the possible absorption from the CN lines and/or from
the \ion{Ca}{I} line, two other synthetic spectra were overlaid: (i)
one spectral synthesis only with the molecular lines and (ii) one
spectral synthesis only with the \ion{Ca}{I}. This visual analysis
also permits to flag or reject some stars from the final Mg
measurement for one of the following reasons: (i) presence of telluric
lines affecting the Mg measurement, (ii) a too low signal to noise
ratio in the Mg region which can affect the continuum placement and
(iii) a strong disagreement between observed and synthetic spectra.   
 
   \begin{figure*}[ht!]
   \centering
   \resizebox{\hsize}{!}{\includegraphics [angle=-90]{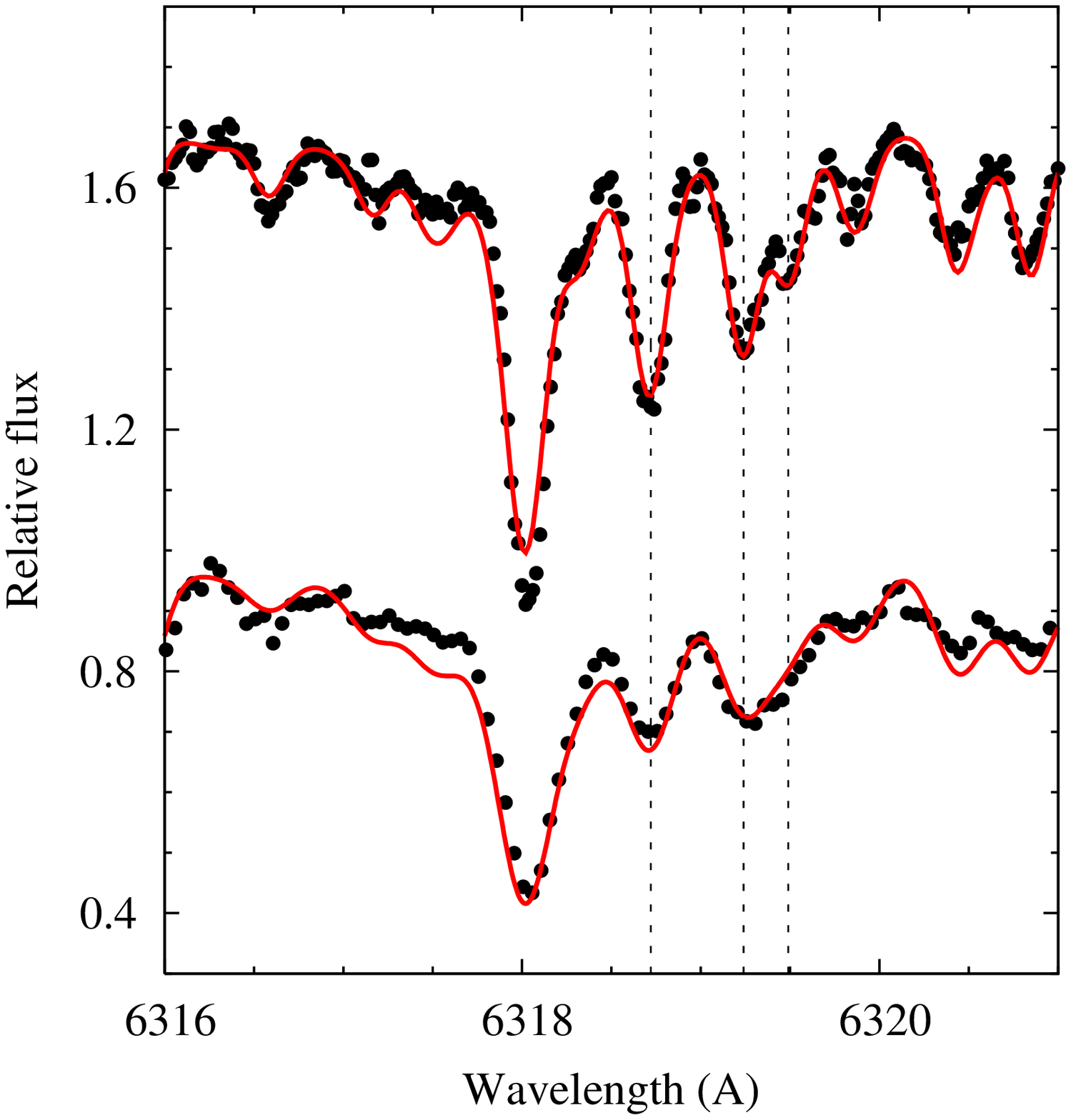}\includegraphics [angle=-90]{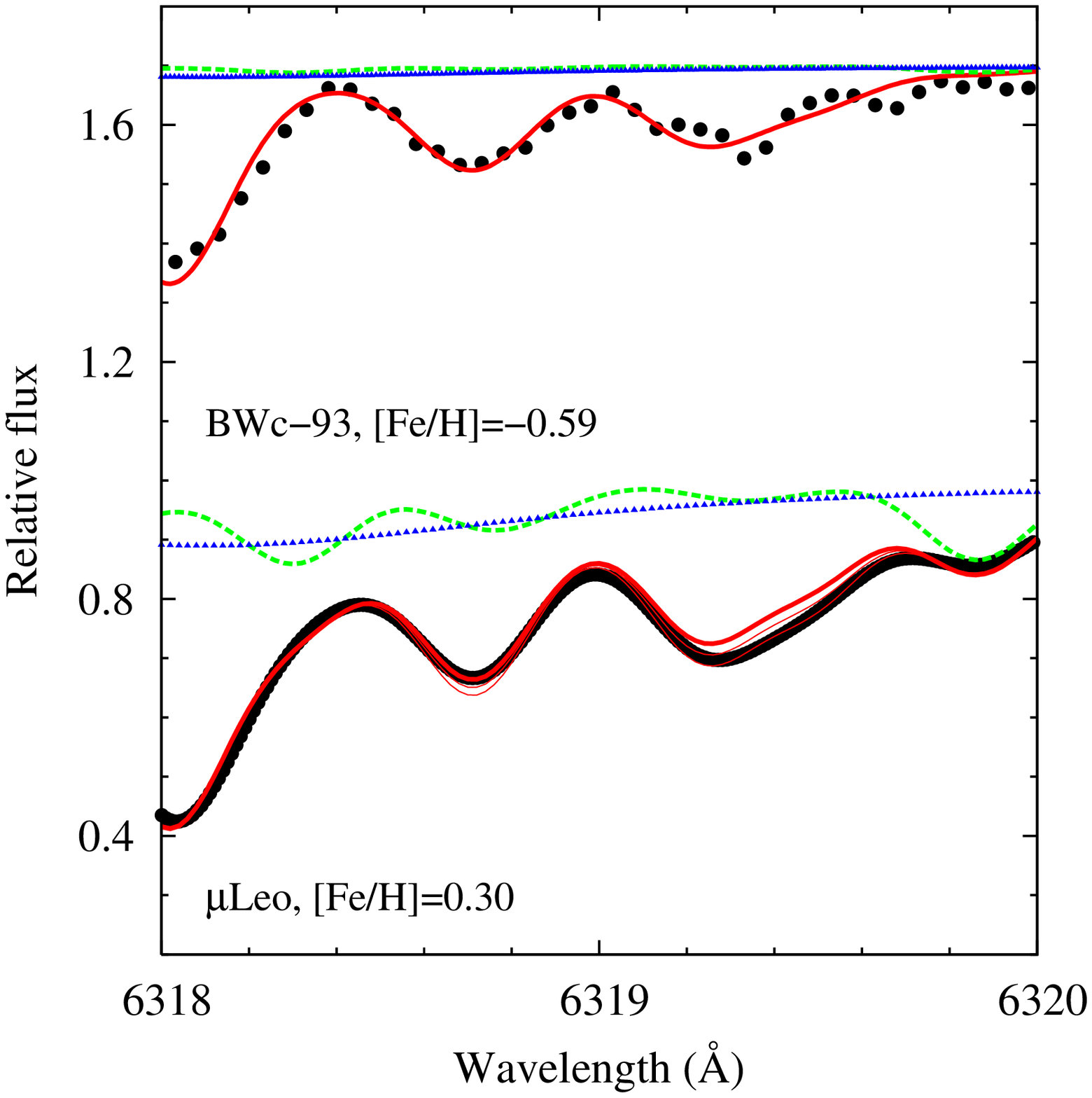}}
    \caption{\textit{Left Panel}: Observed spectrum (black points)
    around the \ion{Mg}{I} triplet for the BWc-1 red clump star
    obtained with GIRAFFE (lower spectrum) and UVES (upper
    spectrum). The red lines represent the synthetic spectra computed
    with [Mg/H]=+0.34 and [Mg/H]=+0.38 respectively for GIRAFFE and UVES
    convolution. \textit{Right Panel}: Comparison of observed (black
    points) and synthetic spectra (red lines) computed with [Mg/H]=-0.16
    for the red clump star BWc-93 and [Mg/H]=+0.42,+0.32,+0.22
    for \MuLeo. The blue dotted lines and the green thick dashed lines
    represent the absorption of the \ion{Ca}{I} line and of the CN
    lines respectively. In BWc-93, the contamination by the
    \ion{Ca}{I} and CN lines is negligible and the agreement between
    the observed and synthetic spectra is found for the three lines of
    the triplet. In \MuLeo, the abundance deduced from the \ion{Mg}{I}
    line at 6318.72 \AA\ is 0.2 dex lower than the one deduced from
    the two redder lines of the triplet.} 
         \label{diff_gir_UVES}
   \end{figure*}

At GIRAFFE resolution, blends between the two redder lines of the
triplet and between the \ion{Mg}{I} line at 6318.72~\AA\ and the
\ion{Fe}{I} line at 6318.1~\AA\ become more important as illustrated by
Fig.\ref{diff_gir_UVES}. This figure also shows that linelist
uncertainties do not have a significant impact on the Mg abundance
determination from UVES spectra, but become larger at GIRAFFE
resolution and can affect the Mg value. More specifically, in the
synthetic spectra, the absorption around 6318.2~\AA\ is overestimated
and at GIRAFFE resolution, the left wing of the \ion{Mg}{I} line at
6318.72~\AA\ become more contaminated leading to an underestimation of
the Mg abundance deduced from this line. These effects were measured
on \MuLeo\ by comparing the observed spectrum convolved to the GIRAFFE
resolution with the synthetic spectrum computed with the stellar
parameters, C, N, O and Ca abundances for \MuLeo\ from
\citet{Lecureur2007} and different Mg abundances (see
Fig.\ref{diff_gir_UVES}, right panel). The abundance deduced from the
two redder lines of the triplet is [Mg/H]=0.42, whereas the abundance
deduced from the line at 6318.72~\AA\ is 0.2~dex lower.  

Due to the previous considerations, for each star of the subsample,
two estimations of the Mg abundance were computed by minimising the
$\rm \chi^{2}$ values between normalised observed and synthetic
spectra on a wavelength domain restricted (i) to the region covered by
the two redder lines ($\rm[Mg/H]_{2Lines}$) and (ii) to the region
covered by the three lines of the Mg triplet
($\rm[Mg/H]_{triplet}$). On the whole sample, the Mg abundance deduced
from the triplet is on the mean 0.05 dex lower than the one deduced
from the two redder lines with a dispersion of 0.06 around the
mean. For the less metallic stars, the difference
$\rm[Mg/H]_{2Lines}-[Mg/H]_{triplet}$ can be negative and the mean
difference is null. In this metallicity range, the triplet absorption
weakens and the abundance deduced from the two redder lines becomes
more sensitive to uncertainties (S/N, continuum placement). For stars
with $\rm[Fe/H]>-0.4$ dex, the difference starts to be positive and
increases with the metallicity of the star to reach 0.15 dex for the
more metallic stars. Moreover, at supersolar metallicities, the
[Mg/Fe] values deduced from the triplet are more dispersed than those
deduced from the two redder lines. In our red clump GIRAFFE sample, to
allow for both these considerations, we finally adopted the [Mg/H]
value deduced from the triplet for the stars with $\rm[Fe/H]<-0.4$ dex
and the value deduced from the two redder lines of the triplet for the
stars with $\rm[Fe/H]>-0.4$ dex. 

\begin{figure}
   \centering
  \resizebox{\hsize}{!}{\includegraphics [angle=-90]{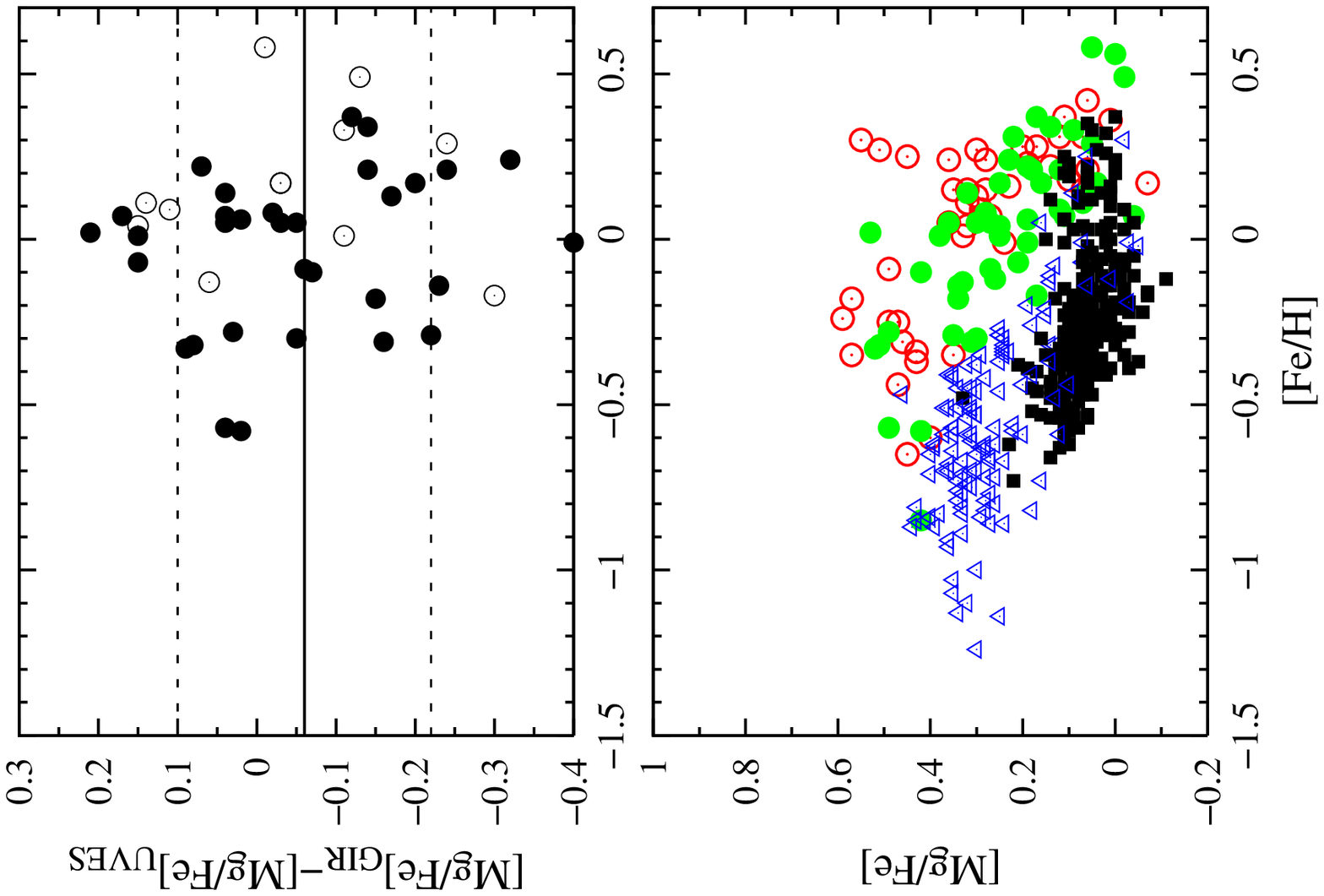}}
    \caption{\textit{Lower Panel}: Comparison of the [Mg/Fe] found
      from GIRAFFE spectra (green filled circles) with [Mg/Fe] found from UVES
      spectra for 44 bulge stars of \citet{Lecureur2007}'s sample 
      (open dotted circles). The thin (black filledÊsquares) and thick 
      (blue open triangles) discs abundances come from
      \cite{Bensby2004,Bensby2005,Reddy2006}. \textit{Upper Panel}:
      Difference on [Mg/Fe] found from GIRAFFE and UVES spectra for 44
      stars (red clump (open circle) and RGB stars (full circle))
      observed with both instruments. The black line shows the mean
      value of this difference (-0.06) and the two dotted lines shows
      the $\rm\pm1\sigma$ around ($\rm\sigma=0.16$). } 
         \label{comp_mg_gir_uves}
   \end{figure}

In order to evaluate the Mg determination for the red clump stars, we
compared the Mg abundances found from the GIRAFFE spectra
($\rm[Mg/H]_{GIR}$) with those found from the UVES spectra
($\rm[Mg/H]_{UVES}$) for the stars observed with both instruments (see
Fig. \ref{comp_mg_gir_uves}). To increase the statistics of the
comparison, we also computed with the same method the Mg abundances in
30 RGB stars of \citet{Lecureur2007}'s sample from the GIRAFFE spectra
with the stellar parameters determined by \cite{Zoccali2008}. For the
total sample (44 stars), a systematic difference between the two
determinations is found: $\rm[Mg/H]_{GIR}-\rm[Mg/H]_{UVES}=-0.06\pm
0.16$ dex which does not depend on the metallicity. This difference is
small, and can be mainly explained by the resolution decrease which
affects the Mg determination via uncertainties in the continuum
placement and/or uncertainties in the atomic and molecular linelist that affect the line fitting differently at different resolutions. Both red clump and RGB stars show the same behaviour.

However, in terms of comparison with abundances found in the discs,
this small difference can become
significant. Fig.~\ref{comp_mg_gir_uves} shows the [Mg/Fe] ratios found
from GIRAFFE and UVES spectra of the same stars. The two sets of
points show a similar global trend, although for $\rm
-0.3<[Fe/H]<0$, $\rm[Mg/Fe]_{GIR}$ ratios are closer to 
those found in thick discs stars at the same metallicity than 
the $\rm[Mg/Fe]_{UVES}$ were. At higher metallicity ($\rm[Fe/H]>0.2$), 
the dispersion on $\rm[Mg/Fe]_{GIR}$ values is considerably lower 
than that of $\rm[Mg/Fe]_{UVES}$.
We attribute the larger dispersion found from 
the UVES spectra to the result of the lower S/N ratios for these spectra.
From both sets of measurements, the bulge stars have [Mg/Fe] ratios on the mean higher than those of the thin disc stars, but half of the stars fall within the thin disc trend from the GIRAFFE measurement, which was not the case from the UVES measurement.
 The [Mg/Fe] trend will be further discussed in Sect. \ref{RAW_DM} for the totalÊred clump sample, but
we would like to note at this point that the previous sample of
\citet{Lecureur2007} plotted here was a mix of red clump (13) in
Baade's Window and RGB stars (44) coming from four different fields
located at Galactic latitudes of $-3^{\circ}$, $-4^{\circ}$ (Baade's
Window), $-6^{\circ}$ to $-12^{\circ}$. Among these fields, some
are expected to be more contaminated by foreground Galactic discs 
(thin and thick), and this
could partly explain the difference between the sample studied in
\citet{Lecureur2007} and the pure Baade's Window sample 
studied in the present paper.  

\subsubsection{[Mg/H] and [Fe/H] uncertainties for the red clump sample}

The individual uncertainties on [Mg/H] and [Fe/H] were estimated from
uncertainties associated with the $\rm \chi^{2}$ procedure ($\rm
\sigma_{FIT}$), as well as the uncertainties linked to uncertainties
on stellar parameters (\Teff\ and \VT\, computed in
Sect.~\ref{stellar_param_error}). 
$\rm \sigma_{FIT}$ was computed using the $\rm \delta \chi^{2}=1$
contour. The uncertainty on [Mg/H] due to the uncertainty on \Teff\
and on \VT\ were computed with the following relations: $\rm
\sigma([Mg/H])_{T}=\left(\frac{d[Mg/H]}{dT}\right)\sigma(T)$ and $\rm
\sigma([Mg/H])_{\xi}=\left(\frac{d[Mg/H]}{d\xi}\right)\sigma(\xi)$
with $\rm\sigma(T)$ and $\rm\sigma(\xi)$ the uncertainties on \Teff\
and \VT\ respectively. $\rm\left(\frac{d[Mg/H]}{dT}\right)$ and
$\rm\left(\frac{d[Mg/H]}{d\xi}\right)$ were estimated from the [Mg/H]
dependance to the \Teff\ and \VT\ variations found in the UVES stars
\citep[see table 10 of][]{Lecureur2007}. Errors arising from log g
uncertainties (an error of 0.3 dex on log g corresponds to an error of
0.02 on average) were found to be negligible compared to the other
sources.

\section{Metallicity distributions}
\subsection{Raw metallicity distributions}\label{RAW_DM}
\subsubsection{[Fe/H] metallicity distribution}

\begin{figure}[!htp]
 \begin{center}
\resizebox{\hsize}{!}{\includegraphics{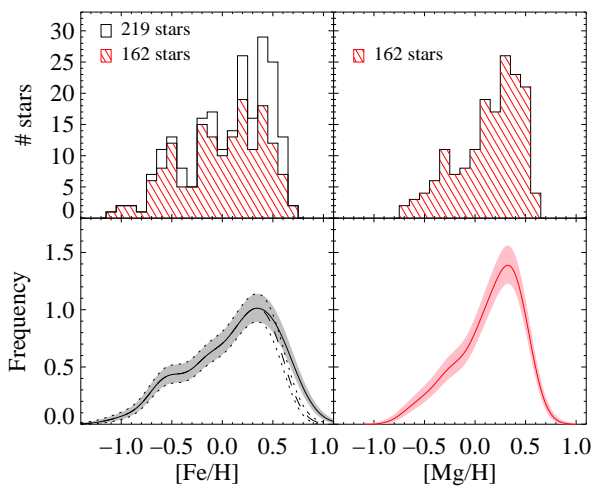}}
       \caption{\textit{Upper left panel:} distribution of [Fe/H] for
       the 219 bulge clump stars in the Baade's window (thick
       histogram) and distribution of [Fe/H] for the 162 star
       subsample for which [Mg/H] is measured (hatched
       histogram). \textit{Upper right panel:} distribution of [Mg/H]
       for the 162 bulge clump stars in the Baade's
       window. \textit{Lower left panel:} smoothed distribution of
       [Fe/H] derived from the upper panel, with its associated
       variability band. The initial metal rich side of the
       distribution (dashed line) has been stretched ((full line) to make it
       consistent with the error law and to achieve the deconvolution
       (see text). \textit{Lower right panel:} 
       smoothed distribution of [Mg/H] derived from the upper panel,
       with its associated variability band (see text).}
\label{MDfig}
 \end{center}
\end{figure}

The resulting metallicity distribution (MD) for the 219 bulge red clump stars is shown in 
Fig.~\ref{MDfig}. The mean [Fe/H] value is $+0.05 \pm 0.03$~dex and the full range of 
metallicities spans $\rm -1.13<[Fe/H]<0.71$ dex. The distribution is clearly asymmetric 
with a median [Fe/H] value of $0.19$~dex and shows a large proportion of metal rich stars: 
65\% with supersolar metallicities and 25\% with $\rm [Fe/H]>0.38$ dex. For $\rm 
[Fe/H]>0.50$ dex, the number of stars decreases considerably. 

The MD shows two low-level secondary peaks around [Fe/H]~=~$\rm-0.55$
and around [Fe/H]~=~$\rm-0.20$ dex, both with very low
statistics. While the first remains whatever the size and centre of
the histogram bins, the second results from a small accumulation of stars in a very narrow metallicity
range around $\rm[Fe/H]=-0.20$ and therefore becomes more or less
pronounced according to binning.  
We checked that this latter peak is not related to the 
two bulge globular clusters NGC6522 and NGC6528 that are close to the region: the former has an intermediate 
metallicity close to $-1$ \citep{Terndrup1998}, while the
latter, with a mean metallicity of $-0.1$ \citep{Zoccali2004} is too
far from the observed window. 

\begin{table}[!htbp]
\begin{center}
\caption{Statistics of the samples:  the size $n$ (number of stars),
  the mean and median (in dex), the smoothing parameter $\hat{h}$
  according to \citet{ShrJos91} and the variability band width
  $\epsilon$.} 
\label{sjeqd}
\begin{tabular}{lrrrrr}
\hline
\hline
\multicolumn{1}{c}{Subsample} & \multicolumn{1}{c}{$n$} &
\multicolumn{1}{c}{mean} & \multicolumn{1}{c}{median} &
\multicolumn{1}{c}{$\hat{h}$}  & \multicolumn{1}{c}{$\epsilon$} \\ 
\hline 
\, [Fe/H] & 219 & $0.05 {\scriptstyle\pm 0.03}$ & $0.16 {\scriptstyle\pm 0.04}$ & 0.0881 & 0.0605\\
\, [Mg/H] & 162 & $0.14 {\scriptstyle\pm 0.02}$ & $0.21 {\scriptstyle\pm 0.03}$ & 0.0871 & 0.0707\\
\hline
\end{tabular}
\end{center}
\end{table}

\label{smoothing}
In order to investigate the significance of low-level secondary peaks,
we estimate the probability density function (PDF) from the raw data,
using a kernel estimator. This estimator is: 
\begin{equation}
\hat\Phi(x) = {1\over n}\sum_{i=1}^{n}{1\over h}\,K\left({x-X_i\over h}\right),
\label{kernelestimator}
\end{equation}
where $(X_i)$ are the observed metallicities, $n$ is the sample size
and $h$ is the window width, also called the smoothing parameter. We
chose the kernel function $K$ to be a Gaussian distribution. The
smoothing parameter $h$ in Eq.~\ref{kernelestimator} is estimated
using the scheme described by \citet{ShrJos91}. The estimator
$\hat{h}$, as solution of the Eq.~12 in \citet{ShrJos91}, is given in
Table~\ref{sjeqd} for each subsample. 

The {\em variability bands}, as described by \citet[ Chap
  2.3]{BonAzi97}, are computed to assess the significance of the modes
in the derived distributions.  
In the case of a Gaussian kernel function, the variability band width
around $\sqrt{\hat\Phi}$ is $\epsilon \approx
{0.2656\,{\left(\hat{h}\,n\right)}^{-0.5}}$ \citep{BonAzi97}. The
values are given in Table~\ref{sjeqd} and they are displayed in the
lower panels in Fig.~\ref{MDfig}. More details about the smoothing
method can be found in \citet[ Sect. 4 and Appendix C]{Ror_07}. 

As far as the secondary modes observed in the [Fe/H] histogram
(Fig.~\ref{MDfig}) are concerned, the smoothed distribution together
with the variability band indicate that the most deficient one, around
$\textrm{[Fe/H]} = -0.55$\,dex, is significant whereas the
intermediate one, around $\textrm{[Fe/H]} = -0.20$\,dex, is not and is
rather part of a transition between the two main modes: around
$\textrm{[Fe/H]} = 0.35$\,dex and $\textrm{[Fe/H]} = -0.55$\,dex. 

\subsubsection{[Mg/H] metallicity distribution}

With the analysis method described in Sect.~\ref{method_measure_Mg}, Mg abundances 
have been determined for a subsample of 162 red clump stars built by
excluding stars with high uncertainties in the [Fe/H] value (only
stars with $\rm \sigma_{up}(Fe)<0.30$ were kept). As illustrated in
Fig. \ref{MDfig}, the resulting [Fe/H] distribution has the same
global shape than the one found before from the total sample, as was
confirmed by a Kolmogorov-Smirnov test (D = 0.0793, p-value =
0.5996). The main difference between the two distributions is for
$\rm[Fe/H]>0.2$ where our selection criteria has excluded a large
number of stars, but the [Fe/H] distribution for the 162 stars still
shows the same sharp decrease of the number of stars at very high
metallicities.  

The resulting [Mg/H] distribution is shown in
Fig. \ref{MDfig}. Compared to the [Fe/H] distribution, the [Mg/H]
distribution is much narrower, ranging from [Mg/H]~=~$\rm-0.7$ to
[Mg/H]~=~$\rm+0.6$. While the sample contains 25\% of stars with
$\rm[Fe/H]<-0.2$, those stars all have $\rm[Mg/Fe]>0$, which explains
the lack of the metal poor tail of the [Mg/H] distribution. The mean
and median values are higher in [Mg/H] than in [Fe/H]: 0.14 and 0.21
respectively (to be compared to 0.05 and 0.16). The distribution also
shows two peaks: one around [Mg/H]~=~$\rm+0.3$ and a second less
significant around $\rm[Mg/H]=-0.3$. For $\rm[MgH]>+0.3$, the number
of stars decreases very abruptly. This characteristic has already been
seen in the [Fe/H] distribution but is even more drastic on the [Mg/H]
distribution. Moreover, we would expect this effect to be larger with
the addition of the stars excluded by the selection. Indeed, the
latter are mainly very high metallicity stars ($\rm[Fe/H]>0.2$) that
should follow the same [Mg/Fe] trend as the other Fe-rich stars (see
Fig.\ref{Mg_sur_Fer}) which have [Mg/Fe] abundances from $-0.1$ to $0.1$
and therefore they should populate the region where
$\rm0.1<[Mg/H]<0.3$. This effect will be discussed further in
Sect. \ref{chemic_evol}. 

The smoothed distribution of [Mg/H] does not display any significant
secondary mode, but the spread of the distribution is indeed smaller
than the one observed for [Fe/H].

\subsubsection{[Fe/H] distribution robustness}\label{bias}

To further investigate the robustness of our iron MD against abundance
undertainties, we rejected stars with high uncertainties in two
different ways, and checked the consistency of the resulting MD: 
{\em  i)} stars with $\rm \sigma_{lo}(Fe)\leqslant0.22$ were kept; 
{\em  ii)} stars with $\rm \sigma_{up}(Fe)\leqslant0.28$ and $\rm
[Fe/H]<0.20$ or  $\rm \sigma_{up}(Fe)\leqslant0.36$ and $\rm
[Fe/H]\geq0.20$  were kept, taking into account the increase of the
$\rm \sigma_{lo}(Fe)$ uncertainty with metallicity (see
Sect.~\ref{stellar_param_error} and Fig.~\ref{errors_fe}). 
The two resulting distributions were found to be fully compatible with the total MD.
Furthermore, to detect possible bias in the MD arising from the
analysis method itself, we checked the robustness of the MD upon the
difference between the spectroscopic and photometric \Teff. The
resulting MD (obtained for the stars with a temperature difference
$\rm\leqslant50$ K and  $\rm\geqslant150$ K) are also compatible with the total MD. 

\begin{figure}[ht!]%
\centering%
  \resizebox{\hsize}{!}{\includegraphics{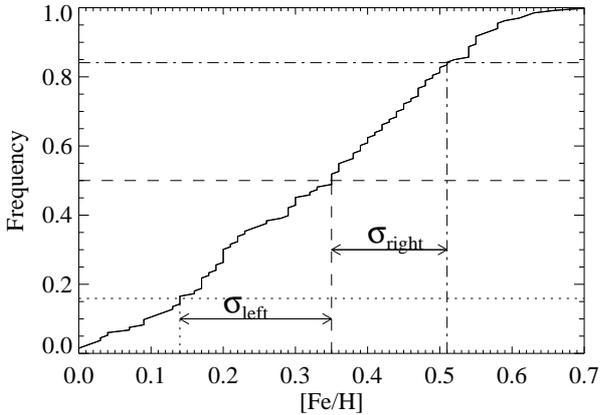}}
\caption{The cumulative iron metallicity distribution for
  $\rm[Fe/H]>0$ (solid line). The median and the left and right
  standard deviations are derived from the frequency values, assuming
  Gaussian errors.} %
\label{cumulrich}
\end{figure}

However, the decrease in the iron MD at the high-metallicity
end (highlighted by the small number of stars with $\rm[Fe/H]>0.5$) is too sharp to be
compatible with the expected measurement errors (see
\ref{stellar_param_error} and Fig.~\ref{errors_fe}) and suggests
that a bias against the highest [Fe/H] could be present.
To show this, let us extract the stars from the MD with
supersolar metallicities, as representative of the metal-rich peak of
the distribution. First, using $\rm[Fe/H]>0$ as a lower cut-off,
the cumulative distribution of supersolar metallicities is plotted
in Fig.~\ref{cumulrich}. Under the hypothesis that, even if the
highest metallicities are biased, the median of the distribution
of the metal-rich peak is
conserved (which is the case if errors remain almost symetrical),
this median (i.e. frequency = 0.5),
corresponds to $\rm [Fe/H]=0.35 dex$. This median is relatively robust to
different cut-off values ($-0.1$ or $0.1$~dex) for defining the high-metallicity peak, for which we obtain
respectively $\rm [Fe/H]=0.32$ and $0.36$~dex. The frequencies $0.16$ and
$0.84$ respectively mark the 1-$\sigma$ positions from the median on the
lower and upper sides of the distribution. These values, displayed in
Fig.~\ref{cumulrich}, show a clear asymetry:
$\sigma_\mathrm{left}=0.21$~dex and
$\sigma_\mathrm{right}=0.16$~dex. While $\sigma_\mathrm{left}$ is
slightly larger than the median expected
measurement error $\rm \sigma_{low,Fe/H]>0}$ of 0.17, allowing for a small astrophysical
scatter, the $\sigma_\mathrm{right}$ is too small compared to the
median expected error
$\rm \sigma_{up,[Fe/H]>0}=0.24$ (Fig.\ref{errors_fe}).
The  metal rich end of the [Fe/H] MD is thus too steep to be
consistent with the expected dispersion of the measurements due to their observational uncertainties only.
One possibility is that remaining degeneracies between $\xi$ $\rm T_{eff}$
and [Fe/H] introduce a bias in the metallicity estimates of
the most metal-rich stars ($\rm [Fe/H]>0.35$) at the resolution a S/N of the present sample, preventing us to truly
establish the underlying shape of the iron MD tail at high
metallicities. 

This bias is not detected in [Mg/H] though, for which the observed [Mg/H]distribution is compatible with the expected uncertainties, even if this distribution also displays a sharp cutoff at high metallicities. This is
because the Mg lines used in the analysis are weak and do not suffer 
in the same way of $\xi$ $\rm T_{eff}$ degeneracies in the stellar parameter
determination. As a result, the uncertainties on [Mg/H] are both smaller than 
those on [Fe/H] at high metallicities, and less assymetric. Another way to 
say this is that, since the error on [Mg/H] is dominated by
actual line measurement errors (synthetic spectra fits in this case),
the possible (small) bias in [Mg/H] at the highest metallicities is
hidden.

To understand wether such a bias could indeed be present, we computed
synthetic spectra of metal-rich (0. to +0.75dex) red clump stars, sliced 
and convolved them to the same wavelength domains, resolution and sampling 
as our observed GIRAFFE spectra. We added photon noise to the spectra to 
reproduce a set of spectra with S/N=100, 50, 40 and 30(100 realisations 
of the noise were generated for each synthetic spectrum and noise level). 
We then retrieved the stellar parameters and metallicity using the same 
method as for the bulge stars (equivalent width measurements and iteration on the stellar parameters). 
The results of these extensive simulations show that
for supersolar metallicities, there is indeed a slight bias in the retrieved
metallicities, although it is in the contrary direction than what would be suggested by the error analysis above: in the mean, very metal-rich stars are found slightly too metal-rich by our method.  In the mean, a [Fe/H]=$+0.5$ star will be found $+0.14$ to $+0.17$\,dex
more metal-rich, whereas a  [Fe/H]=$+0.25$ star will be found $+0.10$ to
$0.14$\,dex more metal-rich (the bias steadily growing with decreasing S/N from
100 to 30). At and below solar metallicities, the bias vanishes.
Finally, we also confirmed this bias by performing the same test on the observed \MuLeo\ spectrum, degraded in resolution, wavelength coverage and photon noise. In this case, the bias is $+0.13$ at S/N=50 and $+0.18$ at S/N=30.

Based on these findings, we warn the reader that there may be a bias in the highest metallicity stars of the sample, although this bias is not understood well enough to be corrected.

\subsubsection{Comparison with the [Fe/H] distribution of RGB stars}\label{SecCompRGB-RC}

\begin{figure}[ht!]
  \resizebox{\hsize}{!}{\includegraphics [angle=-90]{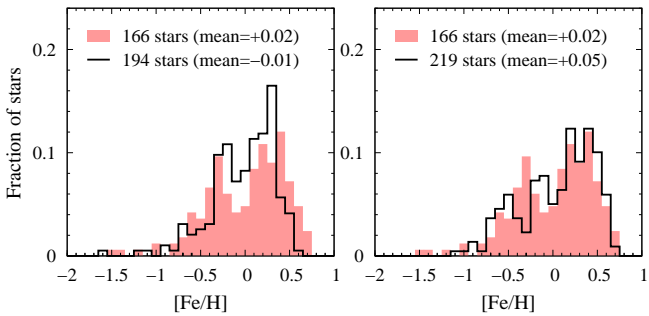}}
\caption{ Iron metallicity distribution (MD) of RGB stars obtained
  with the automatic procedure (red shaded histogram) compared with the MD of
  \cite{Zoccali2008} (\textit{Left panel}, black line) and with the MD
  of the red clump stars (\textit{Right panel}, black line). The new
  RGB MD is in total agreement with the red clump one (\textit{Right
    panel}) but is shifted toward high metallicities compared to the
  MD of \cite{Zoccali2008} (\textit{Left panel}). This small
  difference is due to the difference in analysis of the two samples
  as explained in the text.} %
\label{MD_RGB_and_clump}
\end{figure}

In \cite{Zoccali2008}, another bulge iron MD was obtained in Baade's Window
from a sample of 204 bulge red giants stars. This MD was found in
agreement with the red clump stars one and the small differences
between the two were found to be small enough to combine them to
establish the Baade's Window MD for the comparison with the other
fields (see Sect. 6 of \cite{Zoccali2008}). However, for other
statistical works \citep[see ][]{Babusiaux2010}, the RGB and red clump
MD had to be totally compatible to be combined but a
Kolmogorov-Smirnov (KS) test did not confirm that the 2 samples could
be drawn from a parent population ( p-value = 0.0005). At variance,
the two samples were found to be slightly offsetted one with respect
to the other (the RGB MD was on average 0.07 more metal poor than the
red clump). 

The difference between the two MD could arise from small
differences in the analysis itself. Indeed, even if using the same
criteria to determine the stellar parameters, the analysis of the RGB
samples is different on two points: i) the photometric temperature is
estimated from the V-I index or from the TiO index (whereas it is
computed from V-I, V-J, V-K for the red clump sample) and ii) the
stellar parameters are determined in a manual way (whereas the global
procedure is automatic for the red clump sample). To investigate the
two previous differences, we re-analysed the RGB sample using the same
automatic procedure with two values for the photometric temperature:
i) the same as the one of \cite{Zoccali2008} and ii) a value computed
from the indices V-J, V-H, V-K (as for the red clump sample). 
From the 204 RGB stars sample, we excluded 6 stars as members of the
globular cluster NGC6522 and 4 stars as suspicious binaries and for
the test ii), we also excluded stars for which the 2MASS survey only
gives lower limits in one or more IR bands. Finally, [Fe/H] values
were obtained for 194 stars in the case i) and 166 stars in the case
ii). Before making the comparison between old and new [Fe/H] values,
we checked that the MD of \cite{Zoccali2008} reduced to the 166 stars
of ii) was totally compatible with the total one (204 stars). As shown
by Fig. \ref{MD_RGB_and_clump}, the MD obtained for the 166 stars from
the automatic procedure has a global shape close to the one of
\cite{Zoccali2008} (see Fig. \ref{MD_RGB_and_clump}) with many more
stars at very high metallicities ($\rm[Fe/H]>0.4$) leading to mean and
median values ($0.02$ and $0.11$ respectively) slightly higher than values
previously found ($-0.01$ and $0.05$ respectively). However, in the cases
i) et ii), from the results of a KS test, the MD are not compatible
with the one of \cite{Zoccali2008}: p-value=0.03 for i) and
p-value=0.05 for ii), but they are compatible with the red clump one:
p-value=0.65 for i) and p-value=0.58 for ii) as illustrated by
Fig. \ref{MD_RGB_and_clump} in the case ii). These results show that
the difference between the MD of \cite{Zoccali2008} and the red clump
sample comes from the automatic procedure itself rather than the
initial photometric temperature adopted.  

Note, however, that the difference between the RGB MD presented here
and that of \cite{Zoccali2008} remains very small, and that the conclusions of their
 paper remain fully valid.

\subsection{Unravelling two populations}\label{separ_pop}

\subsubsection{The deconvolved metallicity distribution}

From the PDFs estimated in Sect.~\ref{smoothing} and plotted in
Fig.~\ref{MDfig}, one can rectify the error law effect.
  However, we have argued in Sect.~\ref{bias} that the metal rich end of
  the iron metallicity distribution may be
  biased at the highest metallicities. In order to avoid any
  spurious result in deconvolving  the observed PDF, {we have chosed to strech}Êthe metal-rich side of the MDF, so that $\sigma_\mathrm{right}$ finally
  matches $\sigma_\mathrm{left}$. The stretched distribution is
  superimposed in Fig.~\ref{MDfig}. This stretching will not erase any bias in the highest metallicities if it is indeed present, but helps the deconvolution algorithm to perform nominally.

 We used the Lucy-Richardson deconvolution technique
 \citep{Luy74,Rin72} to recover the rectified MD. More details about
 the deconvolution method can be found in \citet[ Sect. 4 and Appendix
   C]{Ror_07}. 
\par

\begin{figure}[!htp]
 \begin{center}
 \resizebox{\hsize}{!}{\includegraphics{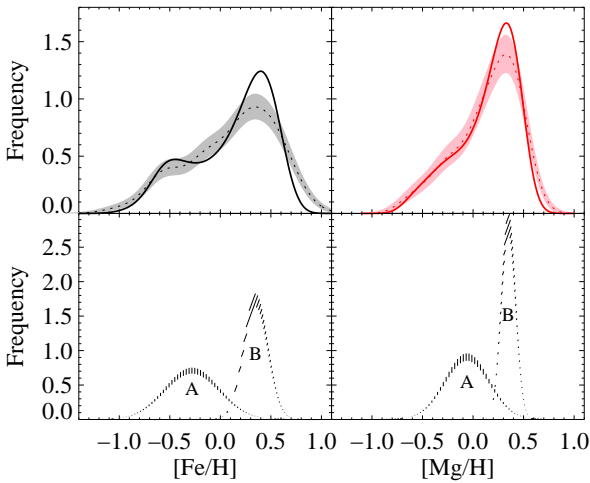}}
       \caption{\textit{Upper left panel:} Smoothed distribution of
       [Fe/H], as in Fig.~\ref{MDfig}, with the resulting deconvolved
       distribution using a constant error (solid line).
       \textit{Upper right panel:} 
       Smoothed distribution of [Mg/H], as in Fig.~\ref{MDfig} with
       the resulting deconvolved distribution using a constant error
       (solid line). \textit{Lower
       left panel:} Resulting Gaussian populations in the [Fe/H]
       distribution. See Table~\ref{gaussfe-mgh} for the parameters of the
       Gaussians. The error bars represent the uncertainty in the
       amplitude of the Gaussian populations.  \textit{Lower right
       panel:} Resulting Gaussian populations in the [Mg/H]
       distribution.}\label{MDfigdeconv}
 \end{center}
\end{figure}

 In this work, the error law is assumed Gaussian.
 We chose to model the error law as a constant, using the median
 of the $(\sigma_\mathrm{low}(\textrm{Fe})+\sigma_\mathrm{up}(\textrm{Fe}))/2$ and
 $\sigma_\mathrm{low}(\textrm{Mg})$: $0.21$\,dex and $0.14$\,dex
 respectively. 
These values are used to estimate the standard deviation in the
Gaussian error distribution, during the Lucy-Richardson
deconvolution. 
The stopping criterion described by \citet{Luy94} was used and the
resulting number of iterations was $11$ and $5$ for the iron and
magnesium distributions respectively. 
The resulting PDF are plotted in Fig.~\ref{MDfigdeconv}.
The deconvolved [Fe/H] distribution shows a clear bimodality with a
low metallicity peak around $-0.45$\,dex and a high metallicity peak
around $+0.3$\,dex. As far as the [Mg/H] is concerned, the
deconvolution does not show clear features in the resulting
distribution. This is due to the fact that [Mg/H] abundances are less
spread and that their associated errors are smaller. 

\subsubsection{Gaussian mixture}
The metallicity distributions described in Section 5.1 suggest that they may be made up
of contributions from different stellar populations. In a first  attempt to identify  
possible sub-structures, we have decomposed the observed [Fe/H] and [Mg/H] distributions into 
a finite number 
of Gaussian components. The SEMMUL algorithm, developed by \cite{Celeux1986}, was
applied. SEMMUL (Stochastic, Estimation, Maximisation, MULtidimensional) is an
iterative method for numerically approximating maximum likelihood estimates for incomplete
data with a stochastic step to accelerate the convergence. It allows separation of Gaussian
components without a priori knowledge of the number of components (only a
maximum number is necessary). In addition, it does require defining a set of
initial conditions. The algorithm has been modified by
\cite{Arenou1993} in order to take observational errors into account, thus the modified
version allows to estimate the cosmic dispersion of the population. 

SEMMUL has been applied to the 1-dimensional [Fe/H] and [Mg/H] distributions separately, and to
the 2-dimensional joint distribution ([Fe/H],[Mg/H]).  Thousands of runs were
carried out in each case, assuming that the
sample is a mixture of 2 or 3 discrete components. 
In all cases, the algorithm converges to a two-component solution
with stable results, dividing the sample in roughly two equal-sized populations.

The resulting two components (that we call A and B) are summarised in Table \ref{gaussfe-mgh}.
For each component, the mean  chemical composition, the dispersion
and the fraction (in percent) of stars in each group, with the
corresponding standard errors, are given. The A and B population
characteristics are very consistent among the three separations
 (one-dimensional along [Fe/H] or [Mg/H], or 2-dimensional). 

In Sect.~\ref{bias} we have argued that the high-metallicity
  tail of the iron distribution may be biased. In this context, we may wonder whether the mean values and
  dispersions obtained in the Gaussian decompositions of population B are
  realistic. The comparison between the estimated mean values 
 ($\rm [Fe/H]=0.32$ and $0.29$, see Table~\ref{gaussfe-mgh}) and the
  median values obtained from the cumulative metallicity distribution
  (for  $\rm[Fe/H]>0$) in Sect.\ref{bias} ($\rm [Fe/H]=0.32$ to $0.36$) shows that they are
  consistent. On the other hand, the obtained cosmic dispersions of
  population B ($0.11$-$0.12$ dex) could arguably be underestimated. 
  However, according to the same cumulative distribution presented in
  Sect.~\ref{bias}, the total metallicity dispersion is about 0.2 dex
  in this population, and the expected measurement errors are of the 
  same order, which implies that the cosmic dispersion of the
  population is small, as obtained by our algorithm.

\subsubsection{Characterising the two populations}

Two populations are clearly identified by the Gaussian  separation
 exercise.
The metal-poor component (A) is centred around $\rm [Fe/H] = -0.30$ and
$\rm [Mg/H] = -0.06$, with a large dispersion ($\sim 0.3$ dex), while the
metal-rich component (B) has similar mean Fe and Mg content ($\rm
 [Fe/H] =0.32$ and $\rm [Mg/H] = +0.35$) and a very small dispersion (0.11 in [Fe/H] and
 0.07 in [Mg/H]). 

The dispersion of population B is very small, comparable in fact
 to the metallicity dispersion of Galactic disc in the solar
 neighbourhood (or even smaller). The more metal-poor population A is
 on the contrary quite extended, both in [Fe/H] and in [Mg/H]. This
 dispersion could in part be due to contamination
 (see Sect.~\ref{section_contam}) by the thick disc in the inner
 Galactic regions, that could  account for $\sim 10-15\%$ of
 population A ($5.9\%$ of the total sample). 
In fact, we will see in the following section that
the chemical signature of this population (A) is not distinct from
that of the thick disc, and we can therefore not exclude that part of
it is made up of thick disc stars, indistinguishable from the
 bona-fide bulge population by any of our observables.

It is further interesting to note that the two components thus
separated are more widely separated in [Fe/H] than in [Mg/H], as
reflected by the different mean corresponding [Mg/Fe] ratios (see
Sect.~\ref{sect_mgfe} and Fig.~\ref{Mg_sur_Fer}), hinting at two physically
separated populations. While the metal-poor component A has a clearly
defined magnesium overabundance of $\rm [Mg/Fe]\sim0.24 $, the
metal-rich component B has a significantly lower $\rm [Mg/Fe] \sim 0.05$, 
not significantly different than the solar neighbourhood
disc magnesium to iron ratio. 

We tentatively explain the existence of these bulge populations as the
result of different origins. In \citet{Babusiaux2010}, we combine the
abundances and the kinematics (radial velocities plus proper motions
from OGLE-II) and show a significant difference in the kinematics of
the metal rich and the metal poor components. The velocity ellipsoid
of the metal rich component shows a vertex deviation consistent with
what is expected from a population kinematically supporting a bar. The
metal poor component on the other hand shows no vertex deviation, 
consistent with an old spheroidal population. 
We therefore relate the richest population to a bar-driven
pseudo-bulge  and the more metal poor one to an old spheroid with a
rapid time-scale formation. Similar results have been suggested by
\cite{Soto2007} and \cite{Zhao1994}, based on kinematics and low resolution metallicities.

Let us note finally that this mixture of
  populations in Baade's Window
is fully compatible with the variation of the metallicity distribution as
a function of Galactic latitude noticed by \citet{Zoccali2008} (from
b=$-4$ to $-12^{\circ}$), that
we would therefore attribute to the gradual disappearance of the
pseudo-bulge component B as one moves away from the Galactic plane,
while the old spheroid (A) would dominate out to higher latitudes. 
The absence of gradients in the inner bulge found by \citet{Rich2007}
  is also consistent, as this region would be dominated by component
  B only.
We also note that population B also has to be predominantly old
  (perhaps not surprisingly if disc formation occurs inside out),
  given that \citet{Clarkson2008} find a pure old population in a
  field at $\rm b=-2.65^{\circ}$ (i.e. significantly closer to the plane
  than Baade's Window).

\begin{table}[htbp]
\begin{center}
\caption{Separation of the sample in two Gaussian components (A and B), with [Fe/H], [Mg/H] 
or both simultaneously as discriminant variables. 
($\rm \sigma_{[Fe/H]}$ and $\rm \sigma_{[Mg/H]}$ are the cosmic dispersion
of the population)}
\label{gaussfe-mgh}
\begin{tabular}{lrrrrr}
\hline\hline
\multicolumn{1}{c}{} & 
\multicolumn{1}{c}{$\rm \overline{[Fe/H]}$} & \multicolumn{1}{c}{$\rm \sigma_{[Fe/H]}$} &
\multicolumn{1}{c}{$\rm \overline{[Mg/H]}$} & \multicolumn{1}{c}{$\rm \sigma_{[Mg/H]}$} & 
\multicolumn{1}{c}{$\%$} \\
\hline
\multicolumn{4}{l}{1D-separation along [Fe/H]}& \multicolumn{2}{r}{$\rm N_{stars} = 219$}\\ 
A & $-0.30 {\scriptstyle \pm 0.03}$ & $0.25 {\scriptstyle \pm 0.01}$ &                  &                 & $45 {\scriptstyle \pm 3}$\\
B & $ 0.32 {\scriptstyle \pm 0.01}$ & $0.11 {\scriptstyle \pm 0.01}$ &                  &                 & $55{\scriptstyle \pm 3}$\\
\hline
\multicolumn{4}{l}{1D-separation along [Mg/H]}& \multicolumn{2}{r}{$\rm N_{stars} = 162$}\\ 
A &                  &                 & $-0.06 {\scriptstyle \pm 0.02}$ & $0.22 {\scriptstyle \pm 0.01}$ & $49 {\scriptstyle \pm 4}$\\
B &                  &                 & $ 0.35 {\scriptstyle \pm 0.01}$ & $0.07 {\scriptstyle \pm 0.01}$ & $51 {\scriptstyle \pm 4}$\\
\hline
\multicolumn{4}{l}{2D-separation in ([Fe/H],[Mg/H])}& \multicolumn{2}{r}{$\rm N_{stars} = 162$}\\ 
A & $-0.27 {\scriptstyle \pm 0.03}$ & $0.31 {\scriptstyle \pm 0.02}$ & $-0.04 {\scriptstyle \pm 0.03}$ & $0.25 {\scriptstyle \pm 0.01}$ & $53 {\scriptstyle \pm 4}$\\
B & $ 0.29 {\scriptstyle \pm 0.01}$ & $0.12 {\scriptstyle \pm 0.01}$ & $ 0.36 {\scriptstyle \pm 0.01}$ & $0.07 {\scriptstyle \pm 0.01}$ & $47 {\scriptstyle \pm 4}$\\
\hline
\end{tabular}
\end{center}
\end{table}

\subsection{The [Mg/Fe] trend in the bulge}\label{sect_mgfe}

\begin{figure*}   
\centering   
\resizebox{\hsize}{!}{\includegraphics {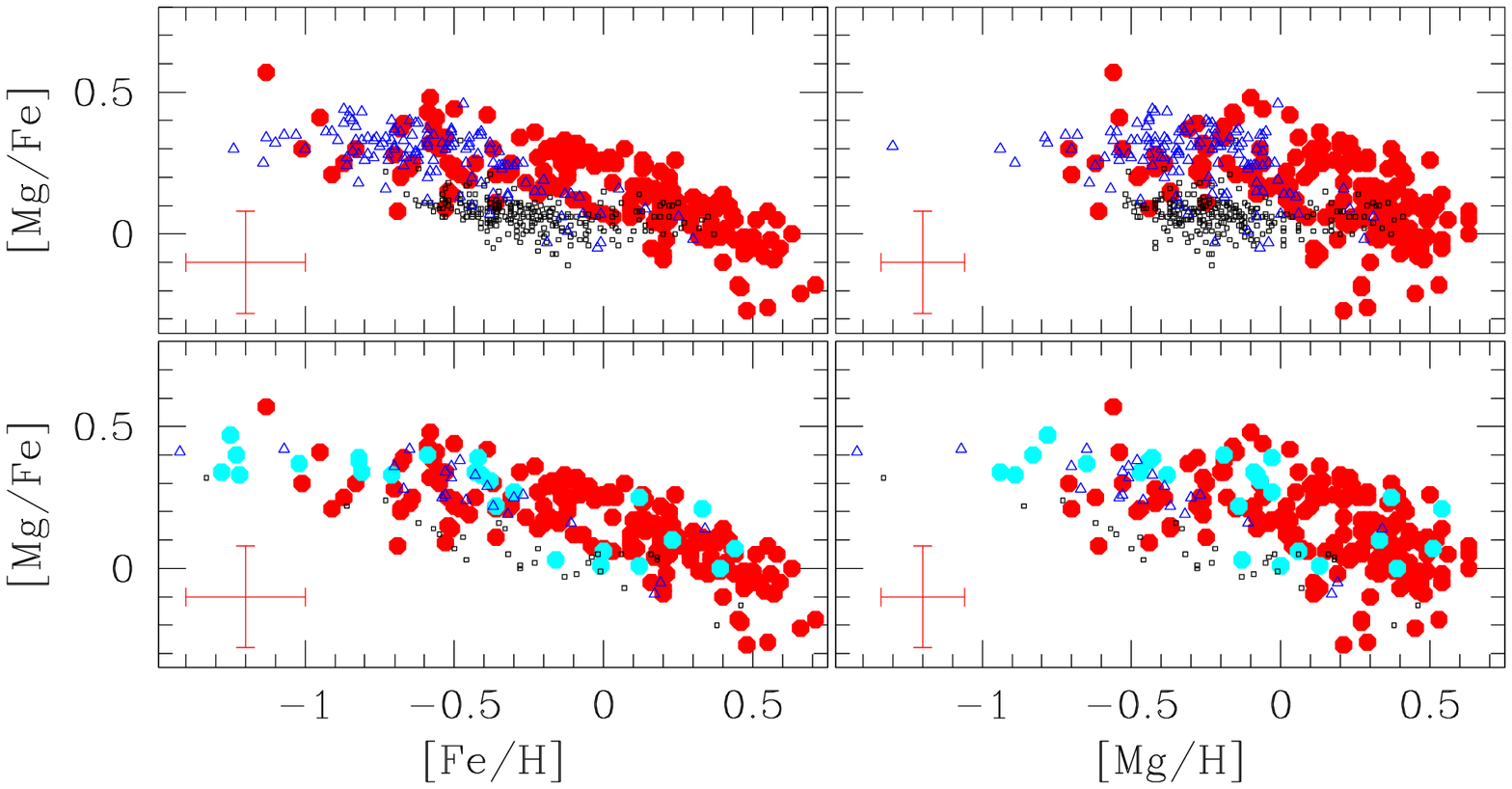}}
  \caption{[Mg/Fe] as a function of [Fe/H] (\textit{right panel}) and [Mg/H]
  (\textit{left panel}) for our bulge red clump sample (red points) compared
  with thick (blue) and thin (black) discs stars. In the two upper panels, the disc reference samples are from the main sequence stars studies of
  \cite{Bensby2005} (triangles), \citet{Reddy2006} (squares)
  \citet{Fuhrmann2008} (crosses), whereas the two lower panels show the 
  solar neighborhood disc giants by \citet{AlvesBrito2010}, as well as their 
  bulge giants (large filled cyan circles). In the low metallicity range
  ($\rm[Fe/H]<-0.3$ and $\rm[Mg/H]<0$), the thick disc and the bulge
  show similar [Mg/Fe] ratios, higher than in thin disc stars.} 
\label{Mg_sur_Fer}
\end{figure*}


Our Mg results in terms of [Mg/Fe] against [Fe/H] and [Mg/H] are shown
in Fig. \ref{Mg_sur_Fer} (red points). The
[Mg/Fe] ratios in the bulge decreases with both [Fe/H] or [Mg/H] from
$\rm\sim+0.5$ to near solar values. Comparing our results with the
recent optical and IR studies of bulge giants from \citet{Rich2005},
\citet{Fulbright2007} and \cite{Rich2007}
on the common metallicity ranges, we note that the general trend is
similar, although our much larger statistics allow us to define the
trend with more confidence.

To understand the relation that the bulge bears to the Galactic
disc(s), it is interesting to compare the [Mg/Fe] ratios in the bulge sample to
the abundance trends found in the Galactic thin and thick discs.
In Fig. \ref{Mg_sur_Fer}, we plotted the thin (black
symbols) and thick (blue symbols) disc stars abundances from the
studies of \cite{Bensby2005,Reddy2006} and \cite{Fuhrmann2008}. These
studies are based on very careful and thoughrough analysis of large
samples of dwarf stars in the solar neighbourhood. As recently pointed out by
\citet{Melendez2008,AlvesBrito2010}, it would be better, for the sake of minimising
possible systematics in the abundance analysis, to compare bulge
giants with solar neighbourhood thin and thick disc giants observed and
analysed in the same conditions. However, from a statistical point of
view, the available samples of local giants available 
for such comparisons is still quite small \citep[ $\sim$20/30 thick/thin disc 
stars respectively, ]{AlvesBrito2010}. 
We therefore decided to stick to a comparison to
local disc dwarfs, once a number of verifications were performed: 
{\it (i)} we analysed the [Mg/Fe] of two reference stars, \MuLeo\ and
Arcturus ([Mg/Fe]=+0.12 and +0.38), that respectively fall on top of
their parent population as derived from local dwarf samples; 
{\it (ii)} \citet{Fulbright2007} analysed a sample of local K giants,
and found no difference between the [Mg/Fe] trend derived from these
giants and those defined from local dwarfs;
{\it (iii)} \citet{Mishenina2006} also analysed a large sample of
local red clump giants, and reach a similar conclusion for [Mg/Fe].
From these checks, we conclude that it is fair to compare the [Mg/Fe] 
abundances for the bulge
red clump giants presented here to those of local disc(s) dwarfs.
In Fig.~\ref{Mg_sur_Fer}, the two lower panels show that indeed 
our conclusions would not be altered in any way if comparing our 
sample to local giants \citep{AlvesBrito2010}.
We however caution that in the highest metallicity regime, systematic
differences of the order of 0.10-0.15 on the [Mg/Fe] trends are found
between different studies of dwarfs stars (for example \cite{Reddy2006}
vs. \cite{Fuhrmann2008}), between dwarf and giant stars samples (for
example \cite{Bensby2005} vs. \cite{Mishenina2006}) and between
different disc giants stars sample (for example \cite{Fulbright2007}
vs \cite{Mishenina2006}). This is clearly a domain where abundances
have to be taken with caution in general, and even more so when they
are derived, as here, from moderate S/N and resolution spectra.

When compared to the local Galactic discs abundances
these various bulge samples show some differences depending on the
metallicity range: 

{\em In the range $\mathit{[Fe/H] \lesssim-0.3}$}, the number of stars in the
present sample has increased dramatically compared to our previous
UVES sample (Fig. \ref{mg_uves}) and they clearly show [Mg/Fe] ratios
similar to those of the thick disc stars for all [Fe/H]. In the same
metallicity range, our results differ from those of
\citet{Fulbright2007} who found higher [Mg/Fe] values than ours and a
[Mg/Fe] trend higher than that of their sample of disc
giants. In fact, if we use Arcturus as a reference to insure that the results of
\citet{Fulbright2007} are on the same scale as the present work,
an offset of $\rm\sim0.10$ is expected
($\rm[Mg/Fe]_{Fulbright}-[Mg/Fe]_{present}=+0.11$), bringing the
[Mg/Fe] of the two studies into agreement.

{\em In the range $\mathit{-0.3\lesssim[Fe/H] \lesssim+0.1}$}, the bulge [Mg/Fe]
trend from our sample is clearly distinct and higher than that of the
thin disc, whatever comparison sample is used. 
In the same metallicity range, our bulge stars show [Mg/Fe]
values on the mean higher than those of the thick disc stars. This
confirms, with better statistics our previous results from the UVES
sample \citep{Lecureur2007} as well as the results of
\citet{Fulbright2007}. Given the residual systematic effect in the
[Mg/Fe] determination  from GIRAFFE and UVES spectra for the
stars in common (see Fig. \ref{diff_gir_UVES}), the difference
between thick disc and bulge would be even more pronounced if we
adopted the UVES abundances as the reference. However, this
difference between bulge and thick disc has to be viewed with
caution because of the small number of thick disc stars in this
metallicity range, and the still debated thick-disc nature of this
high-metallicity tail. Indeed,
\cite{Bensby2005,Bensby2007} argue that the thick disc extends
at least to solar metallicities and shows a clear decrease of
$\rm[\alpha/Fe]$ with increasing iron content for stars with
$\rm [Fe/H]\gtrsim-0.3$ denoting an extended star formation
period (SNIa enrichment). On the contrary, several authors
\citep{Reddy2006,Fuhrmann2008} argue that the thick disc does
not extend significantly in this metallicity regime, and that
there is no evidence of decreasing $\rm[\alpha/Fe]$ in the thick
disc at all.  

{\em In the range $\mathit{[Fe/H] \gtrsim+0.1}$}, the bulge shows [Mg/Fe] ratios
which are similar to those of the local thin disc, solar on the mean,
and with a decreasing [Mg/Fe] trend with increasing metallicity. This
result confirms, with better statistics, the results found for the
oxygen by \cite{Melendez2008} in the same metallicity regime. On the
contrary, our bulge [Mg/Fe] ratios are significantly lower than those
of \citet{Fulbright2007}, even allowing to shift the latter by
$\rm\sim-0.10$ to insure that both studies lie on the same scale. We further note
that \citet{Fulbright2007} derived [Mg/Fe]~$\rm=+0.02$ for \MuLeo,
$\rm0.20$ lower than what we found for this star
\citep{Lecureur2007}. At such high metallicities, both Fe and Mg
measurements remain quite uncertain which could explain part of the
difference between our study and the one of
\citet{Fulbright2007}. Moreover, the comparison of the bulge and disc
trends should be taken with caution in that metallicity regime, since,
as pointed out above, systematic differences of the order of 0.10-0.15
are found between the [Mg/Fe] trends at supersolar metallicities between different
studies. More specifically, taking as a comparison the local disk
sample of giants by \citet{Fulbright2007}, we would find that the
[Mg/Fe] of the bulge lies above that of the local disk, at variance with
the conclusion that is drawn from the comparison to local
dwarfs. However, we do not favour this interpretation since
\citet{Fulbright2007} also finds \MuLeo\ +0.11 dex lower in [Mg/Fe]
than we do.

Very recently, several papers
have measured detailed abundances of a few bulge dwarfs from high
resolution and high signal to noise ratio spectra obtained during a
gravitational microlensing event
\citep{Bensby2010bulge,Bensby2009a,Bensby2009b,Cohen2008,Cohen2009,
Johnson2007,Johnson2008}. 
There are in total by now 14 microlensed dwarfs and subgiants analysed
in the Galactic bulge \citep{Bensby2010bulge}, and although the early (2007 and
2008) microlens events yielded a sample of stars almost exclusively
metal-rich that was irreconcilable with the MDF of bulge giants
\citep{Cohen2008,Cohen2009,Epstein2010}, the 2009 and 2010 events have uncovered a significant metal-poor population, taming down this conclusion to the point that \citet{Bensby2010bulge} claims that the two
distributions could statistically arise from the same parent
population, albeit a rather small probability. The current MDF from these 
microlensed un-evolved stars is now highly bimodal, with a peak at 
$\rm [Fe/H] \sim +0.3$ and another one at $\rm [Fe/H] \sim -0.5$ dex. 
This is very similar to the peaks uncovered by our population separation 
in the red clump sample. Interestingly, the metal-rich (similar to our 
population A) and metal-poor (similar to our population B) microlensed 
stars also have different ages, the latter being a clearly old population, 
while the former spans a range of ages (2-13\,Gyrs). This is again 
compatible with our suggestion that population B is related to the old 
bulge while population A is kinematically related to a bar and would 
be made of a pre-existing disc. 
Even more interestingly perhaps, we are now in a position to compare the
[Mg/Fe] trends in the bulge from dwarf and giants stars: as is our bulge
giants, [Mg/Fe] in dwarfs overlays the local thin disc at super-solar
metallicities, and the local thick disc at metallicities
$[Fe/H]<=-0.3$. The current sample of microlensed giants only contains
two stars in the intermediate metallicity range \citep{Bensby2010bulge},
 where we find the bulge to be richer in [Mg/Fe] than both the local
 thin and thick discs. These two stars seem to trace the upper
 envelope of the local thick disc. To draw any further conclusion on
 this metallicity range, more microlensed dwarfs will need to be analysed.

\begin{figure*}
\centering
\includegraphics[angle=-90,width=\hsize]{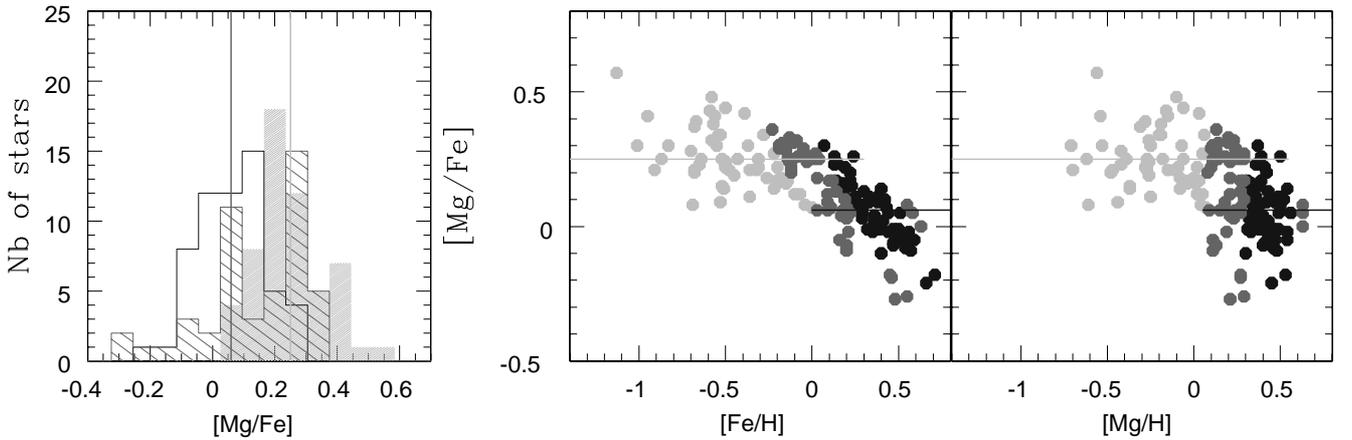}
\caption{\textit{Left panel}: [Mg/Fe] distribution for the stars with
  very high probabilities ($\rm>85$\%) to belong to
  population A (empty histogram) and B (shaded histogram). The two lines show the peaks of these two
  distributions: 0.05 and 0.25 respectively. The hatched histogram
  represents the stars with low probabilities to belong to one or the
  other population. \textit{Right panels}: [Mg/Fe] as a function of
  [Fe/H] and [Mg/Fe] for the three groups of stars.} 
\label{MgFe_separated}
\end{figure*}

To follow up on our tentative sample separation in two distinct
populations we colour-coded in Fig. \ref{MgFe_separated}, the stars
belonging to the populations A and B as separated in
Table~\ref{gaussfe-mgh} along [Mg/H], using the probability with which
a star was assigned to its population. All 
stars with very high probabilities ($> 85\%$) to belong to population A are
coded in light grey, while those with high probability to be in B are
coded in black. Stars that are in dark grey are those with low
probabilities to belong to one or the other class. Population A shows
high [Mg/Fe] ratios, with a mean of $\rm [Mg/Fe]=0.25$, while population B has
lower Mg enhancements, close to solar, as discussed above. 
The intermediate RC population (those stars that could not be unambiguously
classified) has both high and low [Mg/Fe] ratios. In fact, the [Mg/Fe]
histogram of this population shows two well defined peaks, one at $+0.25$
and the other at $+0.05$, coinciding precisely with the means for
population A and B respectively. It therefore seems that this
intermediate population is a mix of stars with the same chemical
properties of A or B population, rather than a smooth transition
between the two. This is also visible in the run of [Mg/Fe] with
metallicity (whether [Fe/H] or [Mg/H], Fig.~\ref{MgFe_separated}) 
at intermediate metallicities, where the stars seem to cluster
around two discrete [Mg/Fe] values extending the trends of population
A and B, rather than follow a smooth decreasing trend. 
As a result, we argue that population A,
that we associate with the old Mg-enhanced bulge, extends up to 
metallicities of at least [Fe/H]=+0.1 and [Mg/H]=+0.35, 
and possibly up to [Fe/H]=+0.25 and [Mg/H]=+0.5. 
Conversely, population B probably extends down to
$\rm [Fe/H]\sim[Mg/H]=0.0$, with modest Mg enhancements. 
This will have an impact on the chemical
evolution models that may be appropriate to represent these two populations.

\section{Chemical evolution and bulge formation}\label{chemic_evol}

 In this section, with the aim of understanding the formation history
 and chemical evolution of the galatic bulge, we compared our bulge MD
 with the results of two recent bulge formation models: the chemical
 evolution model of \citet{Ballero2007} and the chemodynamical model
 of \citet{Immeli2004}.

\subsection{Comparing to a chemical evolution model of the bulge}

\begin{figure}[!htp]
 \begin{center}
 \resizebox{\hsize}{!}{\includegraphics[angle=-90]{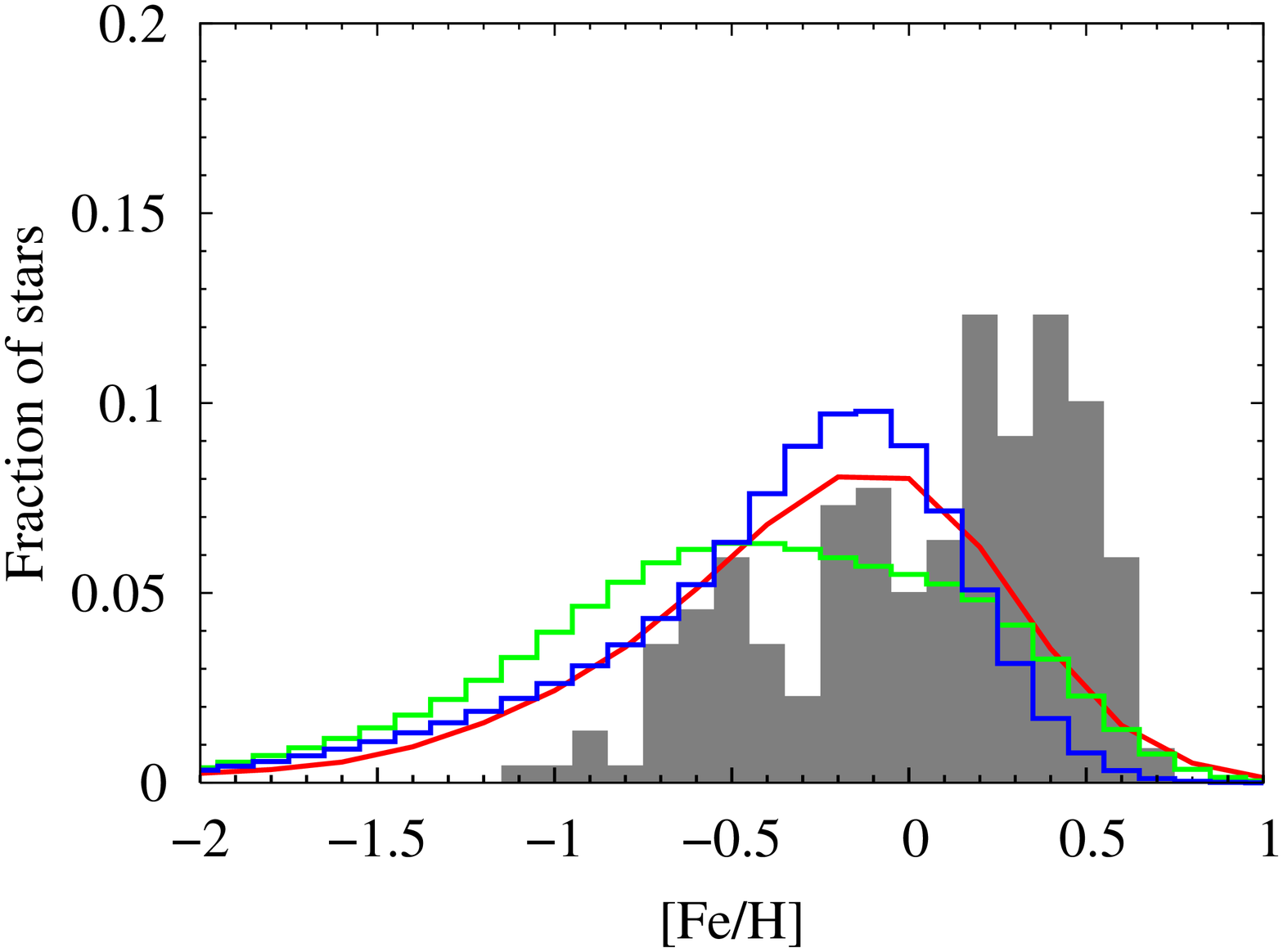}} 
\resizebox{\hsize}{!}{\includegraphics[angle=-90]{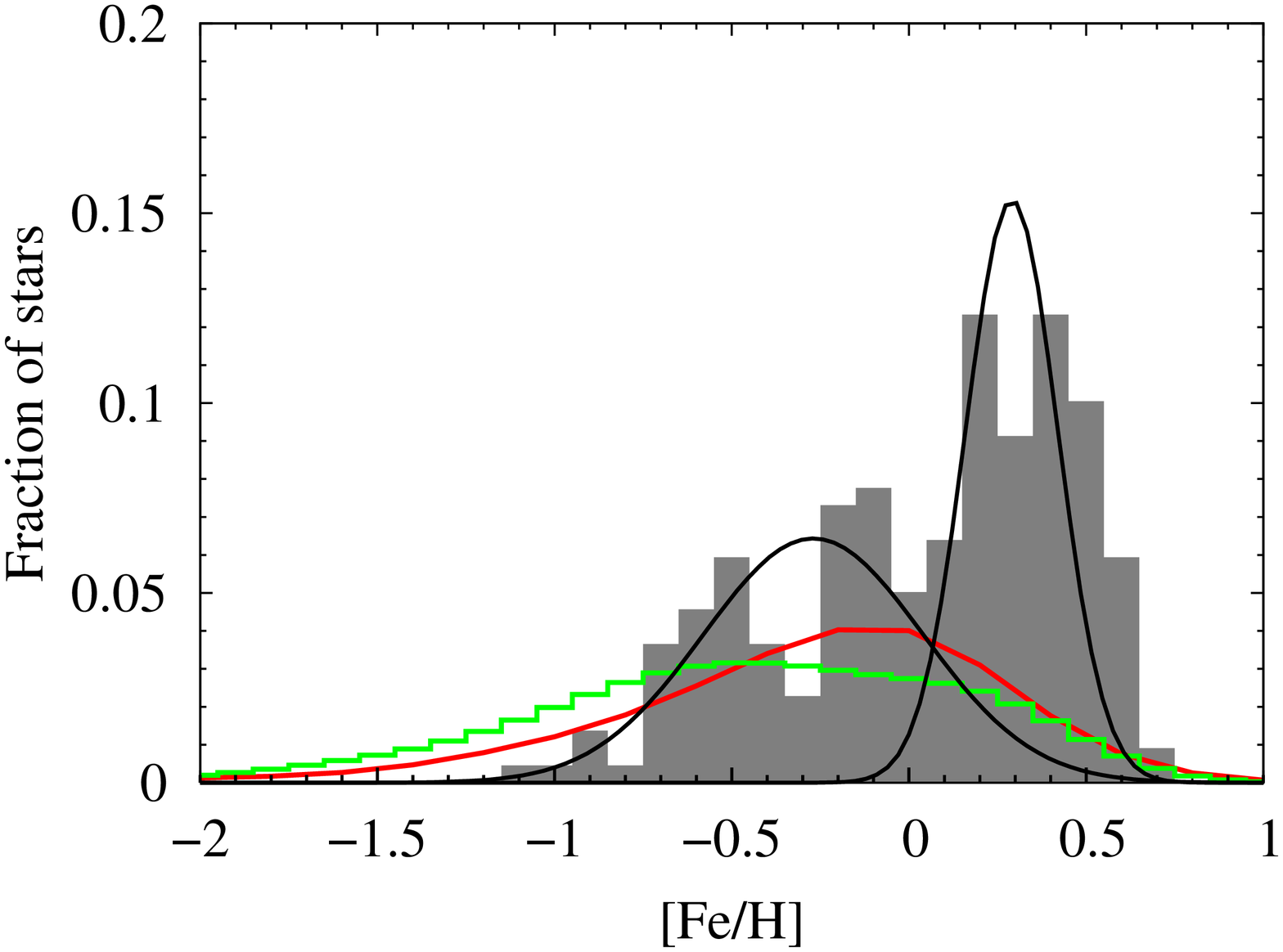}}
       \caption{\textit{Upper panel}. Comparison between the bulge MD
       predicted by the  chemical model of \citet{Ballero2007} (red
       line) and by the model of \citet{Immeli2004} (green line for
       the model B and blue line for the model F ) with the observed
       MD (shaded histogram). The three model MDs have been convolved
       with observational errors. \textit{Lower panel:}. The
       previous  model MDs were normalised to 50\% and compared to the
       MD of the two subpopulations A and B (each of them
       corresponding to 50\% of the total) illustrated here by two
       Gaussians with position and dispersion according to
       Tab. \ref{gaussfe-mgh}.} 
\label{comp_MD_modeles_evol_chim}
 \end{center}
\end{figure}

Figure~\ref{comp_MD_modeles_evol_chim} upper panel shows the MD predicted by the
model of \citet{Ballero2007} together with our observed iron MD. For
the sake of comparison, the model MD has been convolved with the mean
[Fe/H] uncertainties of the observed sample 
(Fig. \ref{errors_fe}). By constraining their model parameters from
bulge observed abundances ratios and Fe MD, \citet{Ballero2007}
concluded that the bulge formed in a very short time, with a high star
formation efficiency ($\rm\mu\simeq20$ Gyr$\rm^{-1}$) and an IMF
flatter than in the solar neighbourhood. The MD they predict peaks
around solar metallicity and is shifted toward lower metallicities with a
more extended metal poor tail with respect to our bulge observed
MD. This disagreement was expected given the differences between our
bulge MD and the one of \citet{Zoccali2003} and \citet{Fulbright2006}
that were used as references by \citet{Ballero2007}. In fact, to
predict near solar metallicities MD consistent with high
$\rm\alpha$-elements ratios, \citet{Ballero2007} had to assume an IMF
flatter than in the solar neighbourhood but also (see
\citet{Ballero2007BIS}) flatter than the stellar IMF of
\citet{Kroupa2001} and the IMF integrated over galaxies of
\citet{Weidner2005}. According to the tests done by the authors, to
shift the predicted MD to more metal-rich (as observed in the present
paper), the IMF should be further flattened. However, for IMF slope
lower than 0.35, the position of the peak of the predicted MD becomes
insensitive to further slope decreases. Moreover, a flatter IMF would
in turn lead to predict higher [Mg/Fe] and [O/Fe] ratios than
observed. Thus, it seems there does not exist any appropriate set of
fundamental parameters for the bulge chemical model of
\citet{Ballero2007} to reproduce both our red clump MD and the [Mg/Fe]
(or [O/Fe]) trend observed in the bulge. 


On the other hand, if we consider the observed MD as the result of two
populations (see Sect. \ref{separ_pop}), it would be more appropriate
to compare the MD of \citet{Ballero2007} (given their bulge chemical
model formation hypothesis) with the one associated to the old bulge
component. This comparison is illustrated by the
Fig. \ref{comp_MD_modeles_evol_chim} where the old bulge population
(A) is modelled by a Gaussian with a mean [Fe/H] of  
-0.3 and a width of 0.25. With its actual parametrisation, the model of
\citet{Ballero2007} also fails to reproduce the metal-poor population
of the bulge. The predicted MD peaks at [Fe/H] values higher than the
mean of population A, and is wider than the Gaussian. It underpredicts
the fraction of stars in population A at intermediate metallicities
($\rm-0.8<[Fe/H]<-0.1$) and predicts a longer low metallicity tail
($\rm [Fe/H<-0.8$). However, based on the parameter variations
  presented by \citet{Ballero2007}, we would expect that it should be
  possible to find a set of parameters that would reproduce both the
  high [Mg/Fe] and metallicity distribution of population A. In
  particular, there would not be the need for a flatter IMF in the
  bulge anymore. 

\subsection{Comparing to a chemodynamical model}

Figure~\ref{comp_MD_modeles_evol_chim} lower panel shows the comparison between our
bulge iron MD and the predicted MD from two models (B and F models) of
\citet{Immeli2004} convolved with the observational errors. 
\citet{Immeli2004} investigated different
scenarios of bulge, disc and halo populations formation from the
evolution of a star-forming disc with a chemodynamical 3D model
including initial conditions from $\rm\Lambda CDM$ results and a two
phases (gas and stars) interstellar medium. The Galactic bulge is
formed in very different way depending on values adopted for the
efficiency of energy dissipation of the cold cloud component from
which stars form. With a high efficiency of energy dissipation, a
massive bulge is formed from the central merging of clumps of stars
and gas at relatively early times (model B). On the contrary, with low
efficiencies values the instability occurs in the stellar disc at
later time forming a stellar bar which evolves and forms a {\em
  pseudo-bulge} (model F).  

The two predicted MD are completely different from the observed
one. As already explained in the previous section, this was expected
for the model B leading to the formation of a massive bulge. However,
the B model MD does not agree either with the metal poor component of
the observed MD (population A). It peaks around [Fe/H]~$\rm=-0.40$,
close to the mean of the Gaussian representative of the old bulge but
is wider than this population, and overproduces the fraction of
metal-poor stars with respect to the actual observed MD.

The observed and model F MDs peak at distinct metallicities
($\rm[Fe/H]\sim+0.3$ and $\rm\sim-0.2$ for the observed and model F
respectively), but have a quite similar global shape. Indeed, if we
artificially shift the model F MD of $\rm+0.5$, the differences
between the two MD are less pronounced, even if differences in the
lowest and highest metallicities regimes persist. The shifted F MD
would overestimate the number of stars at low metallicities
($\rm[Fe/H]<-0.8$ dex) and underestimate the number of stars at high
metallicities ($\rm[Fe/H]>0.1$), in addition to showing a more
progressive decrease of the number of stars than observed at high
metallicities. We will come back in Sect.\ref{Notre_model} on these
differences. However, if we believe the scenario of \citet{Immeli2004}
for the bulge formation (model F), we can not explain why such a large
shift ($\rm+0.5$) in mean metallicity would be required. Uncertainties
in the stellar yields adopted can not explain such a large
difference. To obtain such metal-rich stars, we would suggest that the
gas forming the stellar disc (from which the {\em pseudo-bulge} is
formed) could have been previously enriched. Several factors could
have contributed to the enrichment of this gas, but one possibility
could be enriched gas left over (or blown out) from the old bulge
formation. 

\subsection{An attempt at modelling the two populations}\label{Notre_model}

   \begin{figure*}
   \centering
    \resizebox{\hsize}{!}{\includegraphics[angle=-90]{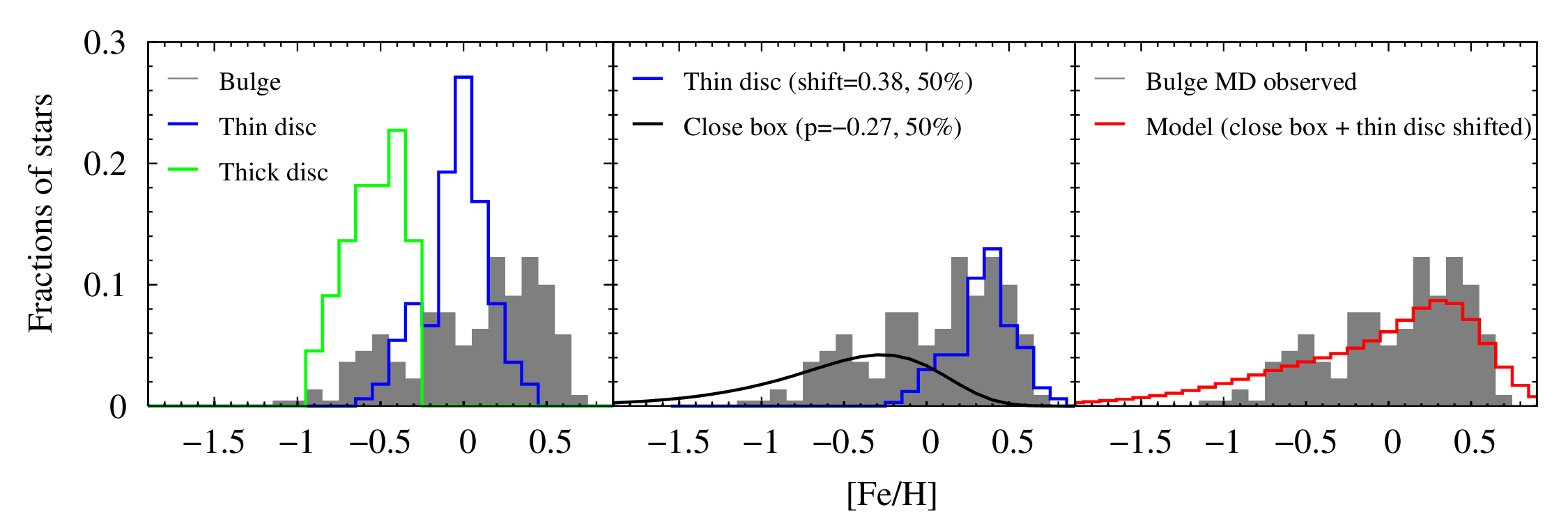}}
  \resizebox{\hsize}{!}{\includegraphics[angle=-90,width=16cm]{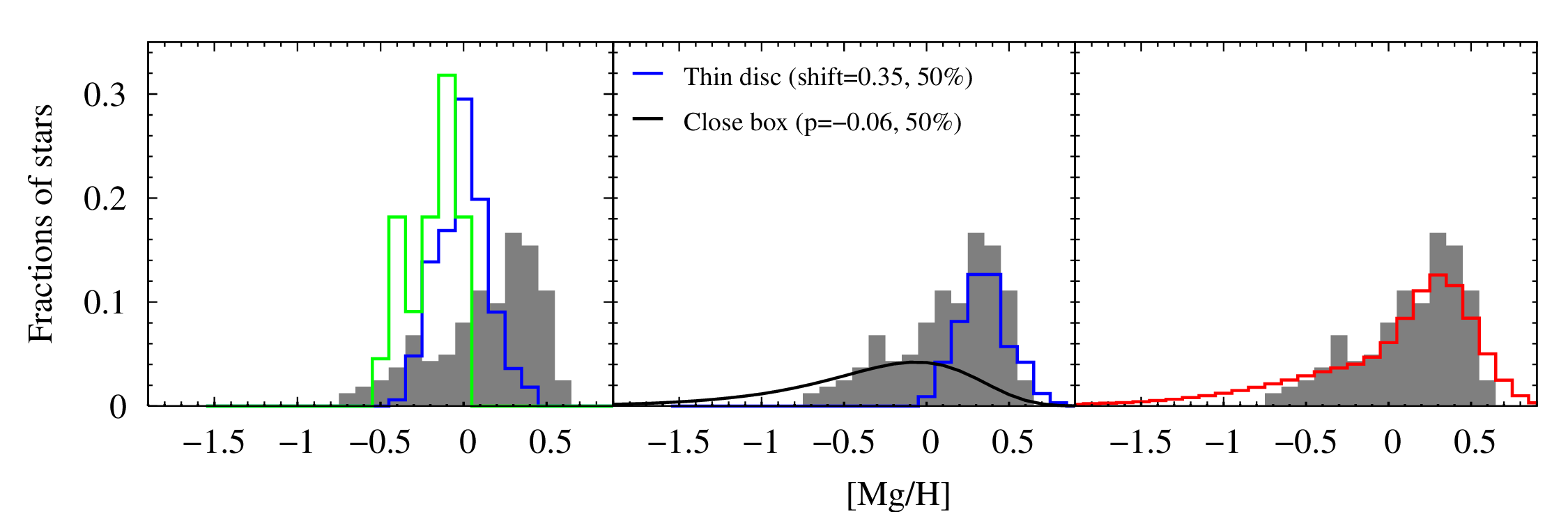}}
    \caption{The \textit{left panels} show the comparison of the
    [Fe/H] MD (\textit{upper}) and [Mg/H] MD (\textit{lower}) with the thin
    (blue) and thick (green) discs MD in the solar neighbourhood from
    the study of \cite{Fuhrmann2008}. The distinction between the
    three populations is clear from their [Fe/H] and [Mg/H] MD in
    their global shapes, their mean values and the metallicity range
    they cover. In the \textit{middle panels} the bulge [Mg/H] and
    [Fe/H] MD are compared to a close box model with a yield $\rm p=-0.27$
    and $-0.06$ for Fe and Mg respectively (black line) and to the
    previous thin disc MD shifted by $+0.35$ (blue) according to the results of the Gaussian
    separation (see Sect. \ref{separ_pop} and Table~\ref{gaussfe-mgh}). The combination of these
    two MD convolved are compared to the observed ones (\textit{right
    panels}) for Fe (\textit{upper}) and Mg (\textit{lower}). For both
    elements, the model MD shows a very good agreement with the
    observed one. A small difference (compared to the relative errors)
    exists on the high metallicities tails: the high metallicity tail
    of the Mg and Fe MD decrease more sharply than the model
    distributions.} 
         \label{COMP_DM_MODEL}
   \end{figure*}

Without any satisfactory bulge formation model that could explain the
observed MD, we tried to build a simple modelling of the two
populations according to their respective formation mechanism
suggested by the results of \citet{Babusiaux2010}.  
The metal-poor population has kinematics that show no influence of the
Galactic bar, and in addition, we find it to be enriched in [Mg/Fe],
which reinforces out interpretation of this population as an old
spheroid with a rapid time-scale formation. To produce the chemical
evolution of this population, we used a simple closed-box model. This
model predicts that the metallicity (Z) distribution function (MDF)
follows the relation $\rm f(Z)=p^{-1}e^{-Z/p}$ where p is the stellar
yield given by $\rm p=<Z>$. For p, we adopted the mean values of the
Gaussian associated to the metal old population: $-0.06$ and 
$-0.30$ [Mg/H] and [Fe/H] respectively. The resulting [Fe/H] and [Mg/H]
metallicities distribution (scaled to 50\% of the total population to
represent population A only)) are compared to the red clump ones on
Fig. \ref{COMP_DM_MODEL} (middle panels).  

As explained in \citet{Babusiaux2010}, the kinematical
 characteristics of the richest population suggest that it is under
 the influence of a bar. Adding to this the fact that [Mg/Fe] in this
 population is near solar, we suggest a formation over a long time
 scale through the evolution of the bar, itself originating in disc
 instability. Within this scenario, population B stars observed in
 Baade's window would have been ejected from the inner regions of the
 Galactic disc. We then expect the chemical composition of these stars
 to be the same as the one of the old inner disc at the time when the
 bar formed. However, as already mentioned, little is known about the
 inner regions of the Galactic disc. A metallicity gradient has been
 observed in young populations (OB stars, HII regions, Cepheids) but
 observational constrains of the old populations are missing. To
 reproduce the MD of the richest population, we adopted the MD of the
 solar neighbourhood from \cite{Fuhrmann2008}, scaled it to 50~\% of
 the total sample (to represent population B only), and shifted toward
 higher metallicity in order to reproduce the possible gradient in the
 Galactic disc. The \cite{Fuhrmann2008} Fe and Mg metallicity
 distributions are symmetric around mean values of  $\rm
 [Fe/H]=-0.03\pm0.01$ and $\rm [Mg/H]=0.00\pm0.01$, and narrower than
 the bulge MD (see  Fig. \ref{COMP_DM_MODEL}, left panels). On the
 middle panels of Fig. \ref{COMP_DM_MODEL}, we plot the
 \cite{Fuhrmann2008} Fe and Mg MD scaled to 50~\% of the total bulge
 population, and shifted by 
 $+0.35$ for both Fe and Mg, so that their mean values coincide with those of the
 corresponding Gaussian components. These Mg and Fe MD shifted disc
 components seem to be good candidates to reproduce the observed bulge
 metal rich component. The main difference (more pronounced for Fe
 than for Mg MD) appears at very high metallicities, where the
 observed bulge MD shows a sharper cutoff than in that of the shifted
 disc. 
 Assuming that this mock disc would originate in the very inner parts
 of the Galactic disc, around 8 kpc from the Sun, the shifts applied
 to the local thin disc MD of \citet{Fuhrmann2008} correspond to
 gradients of $\rm-0.047 dex.kpc^{-1}$ for Fe and $\rm-0.043
 dex.kpc^{-1}$ for Mg. These gradients are similar to the gradients
 that have been observed in the youngest populations of the inner and
 outer disc \citep{Daflon2004}.  
However, in our case we are not interested in the value of the present
 day gradient in the Galactic disc but rather what it was in the past,
 when the stars that we now see in the {\em pseudo-bulge} would have
 formed.  
Recent observations of planetary nebulae and open clusters in the
 Galatic disc \citep[see][ and references therein]{Maciel2009} suggest
 a flattening of the metallicity gradients with time which is not
 consistent with the gradient we derived.  
The chemical evolution models of the disc assuming an inside-out
formation scenario \citep{Cescutti2007,Chiappini2001} predict
gradients at the present day whose values agree with the observations
of young and intermediate-age populations of the disc.  
But, these models also predict that the gradients were flatter at in
the past.  \citet[][ private communication]{Cescutti2007} found that 
9~Gyrs ago, gradients were only of $\rm-0.014 dex.kpc^{-1}$ for Fe and
of $\rm-0.017 dex.kpc^{-1}$ (in the range of galactocentric distances
from 4 to 14 kpc).  
These values, if extended would also be inconsistent with our findings.

On Fig. \ref{COMP_DM_MODEL}, we plotted the sum of the two modelled MD
associated to the metal poor and metal rich components. The MD
predicted by the Simple model have been convolved with the observed
mean uncertainties for our sample (Fig.~\ref{errors_fe}).
The internal errors of the \cite{Fuhrmann2008} data being much  lower than
the observed one, the disc MD have also been convolved with a Gaussian with
$\rm\sigma=0.18$ for Fe and $\rm\sigma=0.12$ for Mg. For both
elements, the combined MD show a fairly good agreement with the observed
ones. The main differences concern the metal poor tail and the richest
part of the distributions.  

In the very metal poor regime ([Fe/H]~$\rm<-0.7$ and
[Mg/Fe]~$\rm<-0.5$), the closed box model predicts more stars with
respect to what is observed. The same lack is observed in the solar
neighbourhood thin disc and is well known as the G dwarfs problem. For
our red clump sample, it is not likely that this difference could come
from a selection effect, at least down to $\rm [Fe/H]=-1.5$ (see
Sect. \ref{section_contam}). As for the local disc, the G dwarf
problem in the bulge could be partly or totally solved by including an
infall of primordial gas in chemical evolution models. Additional
observations of the bulge in the very metal poor regime would also
help to answer this question.  

At higher metallicities, the observed MDs show a peak more pronounced
and a sharper decrease than the shifted MD of the thin disc. These
differences are quasi insensitive to changes up to 0.05 dex on the
adopted shifts for Mg and Fe but could be reduced by decreasing the
adopted width of the convolution (reproducing observed errors) down to
an unrealistic level of 0.10. In Sect.~\ref{bias}, we have argued that
the high-metallicity tail of the iron MD may be biased at the
highest metallicity stars, in a direction that is not yet very clear: experiments with synthetic spectra suggest that we may overestimate the metallicities of super-solar metallicity stars by up to $0.15$\,dex, whereas the estimated errors and sharpness of the decrease of the metal-rich tail may be read as a hint that the highest metallicities are underestimated. 

\section{Summary and conclusions}

We have analysed a sample of 219 bulge red clump stars in Baade's
window from R=20000 resolution spectra obtained with FLAMES/GIRAFFE at
the VLT. This sample has been selected 
based on the location of stars in the Colour-Magnitude Diagram, in
order to minimise the contamination by other Galactic components
($\rm<10$ \% from the latest population synthesis Besan\c{c}on model
\citep{Robin2003,Picaud2004}, and hence is a good tracer of the bulge
metallicity distribution (MD) in the Baade's window. 

For these stars, with an automatic procedure, we have derived the
stellar parameters and Fe abundances using an iron linelist
established differentially to \MuLeo\ and for a subsample of 162 stars
(built by excluding stars with high uncertainties on [Fe/H]), we also
derived the Mg abundances with spectral synthesis around the
\ion{Mg}{I} triplet at $\rm\lambda$ 6319 \AA.  

The Fe and Mg metallicity distributions are both asymmetric, with mean
values of $\rm+0.05 \pm 0.03$ and $\rm+0.14 \pm 0.03$, and median
values of $\rm0.16$ and $\rm0.21$ respectively. They show a small
proportion of stars at low metallicities ($\rm<3$ \% with
$\rm[Fe/H]<-0.7$ or $\rm[Mg/H]<-0.4$), only extending down to
$\rm[Fe/H]=-1.1$ or $\rm[Mg/H]=-0.7$. The iron MD is not well
reproduced by the predictions of the chemical models of the bulge
available in the literature \citep{Ballero2007, Ballero2007BIS},
neither by models adding dynamical aspects to chemical prescriptions
such as the one of \citep{Immeli2004}. 

The decomposition of the observed Fe and Mg MDs into Gaussian
components has revealed the presence of two populations of equal sizes
(50\% each): a metal-poor component centred around [Fe/H]~$\rm=-0.30$
and [Mg/H]~$\rm=-0.06$ with a large dispersion and a narrow metal-rich
component centred around $\rm [Fe/H]=0.32$ and $\rm [Mg/H]=+0.35$. The metal poor
component shows high [Mg/Fe] ratios (around 0.3) whereas stars in the
metal rich component are found to have near solar ratios. 
\citet{Babusiaux2010} also found kinematical differences between the two
components: the metal poor component shows kinematics compatible with
an old spheroid whereas the metal rich component is consistent with a
population supporting a bar. In view of their chemical and kinematical
properties, we suggest different formation scenarios for the two
populations: a rapid formation timescale as an old spheroid for the
metal poor component (old bulge) and for the metal rich component, a
formation over a longer time scale driven by the evolution of the bar
({\em pseudo-bulge}). 
This latter component represents ~50\% of the total Baade's Window 
population, and is of old age \citep[ as indicated by deep colour-magnitude diagrams, see e.g.]{Clarkson2008}, indicating that a fairly massive disc 
was present some 10Gyr ago, which may turn out to be a strong constraint on 
models of disc formation.
It is suggestive that the local thick disc and old bulge display the same
[Mg/Fe] enhancement, pointing at similarly rapid star formation
process. At this point, it is however not possible to constrain further
the relationship that the old bulge bears with the thick disc, both
because of lack of constraints on the 
central parts of the thick disc (both chemistry and kinematics), and
because our bulge sample being located on the minor axis, we have a
very poor constraint on rotation and the disky nature of this
component can therefore not be probed from our data. We however not that in their recent study, \citet{Bensby2010tkd} finds that the inner parts of the thick disc is chemically very similar to that in the solar neighbourhood, lending strong support to the idea that the metal-poor bulge is indeed chemically related to the thick disc.
We also note that the numerous metal-rich globular clusters of the
Milky Way, that are concentrated towards the inner parts of the
galaxy, would be classified as old bulge in our scheme (see e.g. Bica
et al. 2006 that show clusters with $\rm [Fe/H]<-0.75$ are confined
within 2kpc of the centre). It is interesting to note that the most
metal-rich globular cluster known so far is NGC 6528 (Ortolani et
al. 2007), with $\rm [Fe/H]=-0.2$ (Zoccali et al. 2004), therefore within
the metal-poor bulge component identified here.

Guided by these results, we built a simple model combining two
components. We used a simple closed box model to predict the metal
poor population contribution, whereas we represented the metal rich
population by the MD of the local thin disc \citep{Fuhrmann2008},
shifted to the centre of the corresponding metal rich population. The
resulting combined MD shows a good agreement with the red clump one on
the whole metallicity range except for the very metal poor and very
metal rich regimes, where more stars are predicted than observed. The
bulge (old bulge) thus seems to exhibit the G dwarfs problem. 
The  shift we had to apply to the local thin disc MD to fit the
metal-rich population is however quite large (although not impossible)
when compared with plausible values of the gradients in the early
Galactic disc. 
One possibility could be that these stars were born in a pre-enriched
environment, enriched by the remnant of the old bulge. 

The scenario(s) proposed here should now be turned into proper
modelling (chemical and kinematical) of the bulge that would include
two formations events (linked or not), to evaluate their viability.  
Expanding similar combined chemical and kinematical studies of large
samples towards various directions in the Galactic bulge would also be
of great help for answering these questions. In particular, probing
the bulge at lower latitudes and along longitude should trace the
relative importance of the metal-rich and metal-poor populations and
definitively establish their proposed different (kinematical and
chemical) nature as an old bulge and {\em pseudo-bulge}. Because these
regions are much more heavily obscured than Baade's window, this kind
of observations will have to be carried out in the infrared. 
Finally, we note that our group has undertaken similar red clump
observations of the inner parts of the Galactic disc that are expected
to shed new light on the true relation between the disc and the {\em
  pseudo-bulge. }

\begin{acknowledgements}

We thank Bertrand Plez for kindly providing molecular linelists and synthetic spectrum code. 
MZ and DM are supported by FONDAP Center for Astrophysics 15010003, the
BASAL Center for Astrophysics and Associated Technologies PFB-06, the
FONDECYT 1085278 and 1090213, and the MIDEPLAN Milky Way Millennium Nucleus.
\end{acknowledgements}

\bibliographystyle{aa}
\bibliography{biblio}

\begin{thebibliography}{96}
\expandafter\ifx\csname natexlab\endcsname\relax\def\natexlab#1{#1}\fi

\bibitem[{{Aguerri} {et~al.}(2001){Aguerri}, {Balcells}, \&
  {Peletier}}]{Aguerri2001}
{Aguerri}, J.~A.~L., {Balcells}, M., \& {Peletier}, R.~F. 2001, \aap, 367, 428

\bibitem[{{Alonso} {et~al.}(1999){Alonso}, {Arribas}, \&
  {Mart{\'{\i}}nez-Roger}}]{Alonso1999}
{Alonso}, A., {Arribas}, S., \& {Mart{\'{\i}}nez-Roger}, C. 1999, \aaps, 140,
  261

\bibitem[{{Alvarez} \& {Plez}(1998)}]{plezcode}
{Alvarez}, R. \& {Plez}, B. 1998, \aap, 330, 1109

\bibitem[{{Alves-Brito} {et~al.}(2010){Alves-Brito}, {Mel{\'e}ndez}, {Asplund},
  {Ram{\'{\i}}rez}, \& {Yong}}]{AlvesBrito2010}
{Alves-Brito}, A., {Mel{\'e}ndez}, J., {Asplund}, M., {Ram{\'{\i}}rez}, I., \&
  {Yong}, D. 2010, \aap, 513, A35+

\bibitem[{{Arenou}(1993)}]{Arenou1993}
{Arenou}, F. 1993, PhD thesis, Observatoire de Paris, CNRS

\bibitem[{{Babusiaux} \& {Gilmore}(2005)}]{Babusiaux2005}
{Babusiaux}, C. \& {Gilmore}, G. 2005, \mnras, 358, 1309

\bibitem[{{Babusiaux} {et~al.}(2010){Babusiaux}, {Gomez}, {Hill}, {Royer},
  {Zoccali}, \& {Arenou}}]{Babusiaux2010}
{Babusiaux}, C., {Gomez}, A., {Hill}, V., {et~al.} 2010, \aap, 519, A77

\bibitem[{{Ballero} {et~al.}(2007{\natexlab{a}}){Ballero}, {Kroupa}, \&
  {Matteucci}}]{Ballero2007BIS}
{Ballero}, S.~K., {Kroupa}, P., \& {Matteucci}, F. 2007{\natexlab{a}}, \aap,
  467, 117

\bibitem[{{Ballero} {et~al.}(2007{\natexlab{b}}){Ballero}, {Matteucci},
  {Origlia}, \& {Rich}}]{Ballero2007}
{Ballero}, S.~K., {Matteucci}, F., {Origlia}, L., \& {Rich}, R.~M.
  2007{\natexlab{b}}, \aap, 467, 123

\bibitem[{{Bensby} {et~al.}(2010{\natexlab{a}}){Bensby}, {Alves-Brito}, {Oey},
  {Yong}, \& {Mel{\'e}ndez}}]{Bensby2010tkd}
{Bensby}, T., {Alves-Brito}, A., {Oey}, M.~S., {Yong}, D., \& {Mel{\'e}ndez},
  J. 2010{\natexlab{a}}, \aap, 516, L13+

\bibitem[{{Bensby} {et~al.}(2009{\natexlab{a}}){Bensby}, {Feltzing}, {Johnson},
  {Gal-Yam}, {Udalski}, {Gould}, {Han}, {Ad{\'e}n}, \&
  {Simmerer}}]{Bensby2009b}
{Bensby}, T., {Feltzing}, S., {Johnson}, J.~A., {et~al.} 2009{\natexlab{a}},
  \apjl, 699, L174

\bibitem[{{Bensby} {et~al.}(2010{\natexlab{b}}){Bensby}, {Feltzing}, {Johnson},
  {Gould}, {Ad{\'e}n}, {Asplund}, {Mel{\'e}ndez}, {Gal-Yam}, {Lucatello},
  {Sana}, {Sumi}, {Miyake}, {Suzuki}, {Han}, {Bond}, \&
  {Udalski}}]{Bensby2010bulge}
{Bensby}, T., {Feltzing}, S., {Johnson}, J.~A., {et~al.} 2010{\natexlab{b}},
  \aap, 512, A41+

\bibitem[{{Bensby} {et~al.}(2004){Bensby}, {Feltzing}, \&
  {Lundstr{\"o}m}}]{Bensby2004}
{Bensby}, T., {Feltzing}, S., \& {Lundstr{\"o}m}, I. 2004, \aap, 415, 155

\bibitem[{{Bensby} {et~al.}(2005){Bensby}, {Feltzing}, {Lundstr{\"o}m}, \&
  {Ilyin}}]{Bensby2005}
{Bensby}, T., {Feltzing}, S., {Lundstr{\"o}m}, I., \& {Ilyin}, I. 2005, \aap,
  433, 185

\bibitem[{{Bensby} {et~al.}(2009{\natexlab{b}}){Bensby}, {Johnson}, {Cohen},
  {Feltzing}, {Udalski}, {Gould}, {Huang}, {Thompson}, {Simmerer}, \&
  {Ad{\'e}n}}]{Bensby2009a}
{Bensby}, T., {Johnson}, J.~A., {Cohen}, J., {et~al.} 2009{\natexlab{b}}, \aap,
  499, 737

\bibitem[{{Bensby} {et~al.}(2007){Bensby}, {Zenn}, {Oey}, \&
  {Feltzing}}]{Bensby2007}
{Bensby}, T., {Zenn}, A.~R., {Oey}, M.~S., \& {Feltzing}, S. 2007, \apjl, 663,
  L13

\bibitem[{{Blecha} {et~al.}(2000){Blecha}, {Cayatte}, {North}, {Royer}, \&
  {Simond}}]{Blecha2000}
{Blecha}, A., {Cayatte}, V., {North}, P., {Royer}, F., \& {Simond}, G. 2000, in
  Proc. SPIE, Vol. 4008, Optical and IR Telescope Instrumentation and
  Detectors, ed. M.~Iye \& A.~F. Moorwood, 467

\bibitem[{{Bowman} \& {Azzalini}(1997)}]{BonAzi97}
{Bowman}, A.~W. \& {Azzalini}, A. 1997, Applied smoothing techniques for data
  analysis: the kernel approach with S-plus illustrations, Oxford statistical
  science series No.~18 (Clarendon Press)

\bibitem[{{Celeux} \& {Diebolt}(1986)}]{Celeux1986}
{Celeux}, G. \& {Diebolt}, J. 1986, Rev. Stat. Appl., 34, 35

\bibitem[{{Cescutti} {et~al.}(2007){Cescutti}, {Matteucci}, {Fran{\c c}ois}, \&
  {Chiappini}}]{Cescutti2007}
{Cescutti}, G., {Matteucci}, F., {Fran{\c c}ois}, P., \& {Chiappini}, C. 2007,
  \aap, 462, 943

\bibitem[{{Chiappini} {et~al.}(2001){Chiappini}, {Matteucci}, \&
  {Romano}}]{Chiappini2001}
{Chiappini}, C., {Matteucci}, F., \& {Romano}, D. 2001, \apj, 554, 1044

\bibitem[{{Clarkson} {et~al.}(2008){Clarkson}, {Sahu}, {Anderson}, {Smith},
  {Brown}, {Rich}, {Casertano}, {Bond}, {Livio}, {Minniti}, {Panagia},
  {Renzini}, {Valenti}, \& {Zoccali}}]{Clarkson2008}
{Clarkson}, W., {Sahu}, K., {Anderson}, J., {et~al.} 2008, \apj, 684, 1110

\bibitem[{{Cohen} {et~al.}(2008){Cohen}, {Huang}, {Udalski}, {Gould}, \&
  {Johnson}}]{Cohen2008}
{Cohen}, J.~G., {Huang}, W., {Udalski}, A., {Gould}, A., \& {Johnson}, J.~A.
  2008, \apj, 682, 1029

\bibitem[{{Cohen} {et~al.}(2009){Cohen}, {Thompson}, {Sumi}, {Bond}, {Gould},
  {Johnson}, {Huang}, \& {Burley}}]{Cohen2009}
{Cohen}, J.~G., {Thompson}, I.~B., {Sumi}, T., {et~al.} 2009, \apj, 699, 66

\bibitem[{{Combes} \& {Sanders}(1981)}]{Combes1981}
{Combes}, F. \& {Sanders}, R.~H. 1981, \aap, 96, 164

\bibitem[{{Daflon} \& {Cunha}(2004)}]{Daflon2004}
{Daflon}, S. \& {Cunha}, K. 2004, \apj, 617, 1115

\bibitem[{{de Vaucouleurs}(1964)}]{deVaucouleurs1964}
{de Vaucouleurs}, G. 1964, in IAU Symposium, Vol.~20, The Galaxy and the
  Magellanic Clouds, ed. {F.~J.~Kerr}, 195--+

\bibitem[{{Donati} {et~al.}(1997){Donati}, {Semel}, {Carter}, {Rees}, \&
  {Collier Cameron}}]{Donati1997}
{Donati}, J.-F., {Semel}, M., {Carter}, B.~D., {Rees}, D.~E., \& {Collier
  Cameron}, A. 1997, \mnras, 291, 658

\bibitem[{{Dwek} {et~al.}(1995){Dwek}, {Arendt}, {Hauser}, {Kelsall}, {Lisse},
  {Moseley}, {Silverberg}, {Sodroski}, \& {Weiland}}]{Dwek1995}
{Dwek}, E., {Arendt}, R.~G., {Hauser}, M.~G., {et~al.} 1995, \apj, 445, 716

\bibitem[{{Eggen} {et~al.}(1962){Eggen}, {Lynden-Bell}, \&
  {Sandage}}]{Eggen1962}
{Eggen}, O.~J., {Lynden-Bell}, D., \& {Sandage}, A.~R. 1962, \apj, 136, 748

\bibitem[{{Epchtein} {et~al.}(1997){Epchtein}, {de Batz}, {Capoani},
  {Chevallier}, {Copet}, {Fouqu{\'e}}, {Lacombe}, {Le Bertre}, {Pau}, {Rouan},
  {Ruphy}, {Simon}, {Tiph{\`e}ne}, {Burton}, {Bertin}, {Deul}, {Habing},
  {Borsenberger}, {Dennefeld}, {Guglielmo}, {Loup}, {Mamon}, {Ng}, {Omont},
  {Provost}, {Renault}, {Tanguy}, {Kimeswenger}, {Kienel}, {Garzon}, {Persi},
  {Ferrari-Toniolo}, {Robin}, {Paturel}, {Vauglin}, {Forveille}, {Delfosse},
  {Hron}, {Schultheis}, {Appenzeller}, {Wagner}, {Balazs}, {Holl},
  {L{\'e}pine}, {Boscolo}, {Picazzio}, {Duc}, \& {Mennessier}}]{DENIS_survey}
{Epchtein}, N., {de Batz}, B., {Capoani}, L., {et~al.} 1997, The Messenger, 87,
  27

\bibitem[{{Epstein} {et~al.}(2010){Epstein}, {Johnson}, {Dong}, {Udalski},
  {Gould}, \& {Becker}}]{Epstein2010}
{Epstein}, C.~R., {Johnson}, J.~A., {Dong}, S., {et~al.} 2010, \apj, 709, 447

\bibitem[{{Feltzing} \& {Gilmore}(2000)}]{Feltzing2000}
{Feltzing}, S. \& {Gilmore}, G. 2000, \aap, 355, 949

\bibitem[{{Fuhrmann}(2008)}]{Fuhrmann2008}
{Fuhrmann}, K. 2008, \mnras, 384, 173

\bibitem[{{Fulbright} {et~al.}(2006){Fulbright}, {McWilliam}, \&
  {Rich}}]{Fulbright2006}
{Fulbright}, J.~P., {McWilliam}, A., \& {Rich}, R.~M. 2006, \apj, 636, 821

\bibitem[{{Fulbright} {et~al.}(2007){Fulbright}, {McWilliam}, \&
  {Rich}}]{Fulbright2007}
{Fulbright}, J.~P., {McWilliam}, A., \& {Rich}, R.~M. 2007, \apj, 661, 1152

\bibitem[{{Gerhard}(2006)}]{Gerhard2006}
{Gerhard}, O. 2006, in EAS Publications Series, Vol.~20, EAS Publications
  Series, ed. G.~A. {Mamon}, F.~{Combes}, C.~{Deffayet}, \& B.~{Fort}, 89--96

\bibitem[{{Girardi} {et~al.}(2000){Girardi}, {Bressan}, {Bertelli}, \&
  {Chiosi}}]{Girardi2000}
{Girardi}, L., {Bressan}, A., {Bertelli}, G., \& {Chiosi}, C. 2000, \aaps, 141,
  371

\bibitem[{{Glass}(1999)}]{Glass1999}
{Glass}, I.~S. 1999, {Handbook of Infrared Astronomy} (Cambridge University
  Press)

\bibitem[{{Gustafsson} {et~al.}(2008){Gustafsson}, {Edvardsson}, {Eriksson},
  {J{\o}rgensen}, {Nordlund}, \& {Plez}}]{OSMARCS}
{Gustafsson}, B., {Edvardsson}, B., {Eriksson}, K., {et~al.} 2008, \aap, 486,
  951

\bibitem[{{Howard} {et~al.}(2009){Howard}, {Rich}, {Clarkson}, {Mallery},
  {Kormendy}, {De Propris}, {Robin}, {Fux}, {Reitzel}, {Zhao}, {Kuijken}, \&
  {Koch}}]{Howard2009}
{Howard}, C.~D., {Rich}, R.~M., {Clarkson}, W., {et~al.} 2009, \apjl, 702, L153

\bibitem[{{Howard} {et~al.}(2008){Howard}, {Rich}, {Reitzel}, {Koch}, {De
  Propris}, \& {Zhao}}]{Howard2008}
{Howard}, C.~D., {Rich}, R.~M., {Reitzel}, D.~B., {et~al.} 2008, \apj, 688,
  1060

\bibitem[{{Ibata} \& {Gilmore}(1995{\natexlab{a}})}]{Ibata1995a}
{Ibata}, R.~A. \& {Gilmore}, G.~F. 1995{\natexlab{a}}, \mnras, 275, 591

\bibitem[{{Ibata} \& {Gilmore}(1995{\natexlab{b}})}]{Ibata1995b}
{Ibata}, R.~A. \& {Gilmore}, G.~F. 1995{\natexlab{b}}, \mnras, 275, 605

\bibitem[{{Immeli} {et~al.}(2004){Immeli}, {Samland}, {Gerhard}, \&
  {Westera}}]{Immeli2004}
{Immeli}, A., {Samland}, M., {Gerhard}, O., \& {Westera}, P. 2004, \aap, 413,
  547

\bibitem[{{Johnson} {et~al.}(2007){Johnson}, {Gal-Yam}, {Leonard}, {Simon},
  {Udalski}, \& {Gould}}]{Johnson2007}
{Johnson}, J.~A., {Gal-Yam}, A., {Leonard}, D.~C., {et~al.} 2007, \apjl, 655,
  L33

\bibitem[{{Johnson} {et~al.}(2008){Johnson}, {Gaudi}, {Sumi}, {Bond}, \&
  {Gould}}]{Johnson2008}
{Johnson}, J.~A., {Gaudi}, B.~S., {Sumi}, T., {Bond}, I.~A., \& {Gould}, A.
  2008, \apj, 685, 508

\bibitem[{{Kormendy} \& {Kennicutt}(2004)}]{Kormendy2004}
{Kormendy}, J. \& {Kennicutt}, Jr., R.~C. 2004, \araa, 42, 603

\bibitem[{{Kroupa}(2001)}]{Kroupa2001}
{Kroupa}, P. 2001, \mnras, 322, 231

\bibitem[{{Kuijken} \& {Rich}(2002)}]{Kuijken2002}
{Kuijken}, K. \& {Rich}, R.~M. 2002, \aj, 124, 2054

\bibitem[{{Kupka} {et~al.}(1999){Kupka}, {Piskunov}, {Ryabchikova}, {Stempels},
  \& {Weiss}}]{Kupka1999}
{Kupka}, F., {Piskunov}, N., {Ryabchikova}, T.~A., {Stempels}, H.~C., \&
  {Weiss}, W.~W. 1999, \aaps, 138, 119

\bibitem[{{Lecureur} {et~al.}(2007){Lecureur}, {Hill}, {Zoccali}, {Barbuy},
  {G{\'o}mez}, {Minniti}, {Ortolani}, \& {Renzini}}]{Lecureur2007}
{Lecureur}, A., {Hill}, V., {Zoccali}, M., {et~al.} 2007, \aap, 465, 799

\bibitem[{{Lucy}(1974)}]{Luy74}
{Lucy}, L.~B. 1974, \aj, 79, 745

\bibitem[{{Lucy}(1994)}]{Luy94}
{Lucy}, L.~B. 1994, \aap, 289, 983

\bibitem[{{Maciel} \& {Costa}(2009)}]{Maciel2009}
{Maciel}, W.~J. \& {Costa}, R.~D.~D. 2009, in IAU Symposium, Vol. 254, IAU
  Symposium, ed. {J.~Andersen, J.~Bland-Hawthorn, \& B.~Nordstr{\"o}m}, 38P--+

\bibitem[{{Magain}(1984)}]{Magain1984}
{Magain}, P. 1984, \aap, 134, 189

\bibitem[{{McWilliam} {et~al.}(1995){McWilliam}, {Preston}, {Sneden}, \&
  {Searle}}]{McWilliam1995}
{McWilliam}, A., {Preston}, G.~W., {Sneden}, C., \& {Searle}, L. 1995, \aj,
  109, 2757

\bibitem[{{McWilliam} \& {Rich}(1994)}]{McWilliam1994}
{McWilliam}, A. \& {Rich}, R.~M. 1994, \apjs, 91, 749

\bibitem[{{Mel{\'e}ndez} {et~al.}(2008){Mel{\'e}ndez}, {Asplund},
  {Alves-Brito}, {Cunha}, {Barbuy}, {Bessell}, {Chiappini}, {Freeman},
  {Ram{\'{\i}}rez}, {Smith}, \& {Yong}}]{Melendez2008}
{Mel{\'e}ndez}, J., {Asplund}, M., {Alves-Brito}, A., {et~al.} 2008, \aap, 484,
  L21

\bibitem[{{Minniti}(1996)}]{Minniti1996}
{Minniti}, D. 1996, \apj, 459, 175

\bibitem[{{Mishenina} {et~al.}(2006){Mishenina}, {Bienaym{\'e}}, {Gorbaneva},
  {Charbonnel}, {Soubiran}, {Korotin}, \& {Kovtyukh}}]{Mishenina2006}
{Mishenina}, T.~V., {Bienaym{\'e}}, O., {Gorbaneva}, T.~I., {et~al.} 2006,
  \aap, 456, 1109

\bibitem[{{Nakasato} \& {Nomoto}(2003)}]{Nakasato2003}
{Nakasato}, N. \& {Nomoto}, K. 2003, \apj, 588, 842

\bibitem[{{Nishiyama} {et~al.}(2006){Nishiyama}, {Nagata}, {Sato}, {Kato},
  {Nagayama}, {Kusakabe}, {Matsunaga}, {Naoi}, {Sugitani}, \&
  {Tamura}}]{Nishiyama2006}
{Nishiyama}, S., {Nagata}, T., {Sato}, S., {et~al.} 2006, \apj, 647, 1093

\bibitem[{{Noguchi}(1999)}]{Noguchi1999}
{Noguchi}, M. 1999, \apj, 514, 77

\bibitem[{{Ortolani} {et~al.}(1995){Ortolani}, {Renzini}, {Gilmozzi},
  {Marconi}, {Barbuy}, {Bica}, \& {Rich}}]{Ortolani1995}
{Ortolani}, S., {Renzini}, A., {Gilmozzi}, R., {et~al.} 1995, \nat, 377, 701

\bibitem[{{Paczynski} {et~al.}(1999){Paczynski}, {Udalski}, {Szymanski},
  {Kubiak}, {Pietrzynski}, {Soszynski}, {Wozniak}, \& {Zebrun}}]{Paczynski1999}
{Paczynski}, B., {Udalski}, A., {Szymanski}, M., {et~al.} 1999, Acta
  Astronomica, 49, 319

\bibitem[{{Pfenniger} \& {Norman}(1990)}]{Pfenniger1990}
{Pfenniger}, D. \& {Norman}, C. 1990, \apj, 363, 391

\bibitem[{{Picaud} \& {Robin}(2004)}]{Picaud2004}
{Picaud}, S. \& {Robin}, A.~C. 2004, \aap, 428, 891

\bibitem[{{Plez}(1998)}]{Plez1998}
{Plez}, B. 1998, \aap, 337, 495

\bibitem[{{Raha} {et~al.}(1991){Raha}, {Sellwood}, {James}, \&
  {Kahn}}]{Raha1991}
{Raha}, N., {Sellwood}, J.~A., {James}, R.~A., \& {Kahn}, F.~D. 1991, \nat,
  352, 411

\bibitem[{{Ralchenko} {et~al.}(2008){Ralchenko}, {Kramida}, {Reader}, \& {NIST
  ASD Team}}]{NIST}
{Ralchenko}, Y., {Kramida}, A.~E., {Reader}, J., \& {NIST ASD Team}. 2008,
  Available: http://physics.nist.gov/asd3, National Institute of Standards and
  Technology, Gaithersburg, MD.

\bibitem[{{Ram{\'{\i}}rez} \& {Mel{\'e}ndez}(2005)}]{Ramirez2005}
{Ram{\'{\i}}rez}, I. \& {Mel{\'e}ndez}, J. 2005, \apj, 626, 465

\bibitem[{{Rattenbury} {et~al.}(2007){Rattenbury}, {Mao}, {Sumi}, \&
  {Smith}}]{Rattenbury2007a}
{Rattenbury}, N.~J., {Mao}, S., {Sumi}, T., \& {Smith}, M.~C. 2007, \mnras,
  378, 1064

\bibitem[{{Reddy} {et~al.}(2006){Reddy}, {Lambert}, \& {Allende
  Prieto}}]{Reddy2006}
{Reddy}, B.~E., {Lambert}, D.~L., \& {Allende Prieto}, C. 2006, \mnras, 367,
  1329

\bibitem[{{Reid}(1993)}]{Reid1993}
{Reid}, M.~J. 1993, \araa, 31, 345

\bibitem[{{Rich}(1988)}]{Rich1988}
{Rich}, R.~M. 1988, \aj, 95, 828

\bibitem[{{Rich} \& {Origlia}(2005)}]{Rich2005}
{Rich}, R.~M. \& {Origlia}, L. 2005, \apj, 634, 1293

\bibitem[{{Rich} {et~al.}(2007){Rich}, {Origlia}, \& {Valenti}}]{Rich2007}
{Rich}, R.~M., {Origlia}, L., \& {Valenti}, E. 2007, \apjl, 665, L119

\bibitem[{{Richardson}(1972)}]{Rin72}
{Richardson}, W.~H. 1972, J. Opt. Soc. Am., 62, 55

\bibitem[{{Robin} {et~al.}(2003){Robin}, {Reyl{\'e}}, {Derri{\`e}re}, \&
  {Picaud}}]{Robin2003}
{Robin}, A.~C., {Reyl{\'e}}, C., {Derri{\`e}re}, S., \& {Picaud}, S. 2003,
  \aap, 409, 523

\bibitem[{{Royer} {et~al.}(2007){Royer}, {Zorec}, \& {G{\'o}mez}}]{Ror_07}
{Royer}, F., {Zorec}, J., \& {G{\'o}mez}, A.~E. 2007, \aap, 463, 671

\bibitem[{{Sadler} {et~al.}(1996){Sadler}, {Rich}, \& {Terndrup}}]{Sadler1996}
{Sadler}, E.~M., {Rich}, R.~M., \& {Terndrup}, D.~M. 1996, \aj, 112, 171

\bibitem[{{Scannapieco} \& {Tissera}(2003)}]{Scannapieco2003}
{Scannapieco}, C. \& {Tissera}, P.~B. 2003, \mnras, 338, 880

\bibitem[{{Sheather} \& {Jones}(1991)}]{ShrJos91}
{Sheather}, S.~J. \& {Jones}, M.~C. 1991, J. R. Statist. Soc B., 53, 683

\bibitem[{{Skrutskie} {et~al.}(2006){Skrutskie}, {Cutri}, {Stiening},
  {Weinberg}, {Schneider}, {Carpenter}, {Beichman}, {Capps}, {Chester},
  {Elias}, {Huchra}, {Liebert}, {Lonsdale}, {Monet}, {Price}, {Seitzer},
  {Jarrett}, {Kirkpatrick}, {Gizis}, {Howard}, {Evans}, {Fowler}, {Fullmer},
  {Hurt}, {Light}, {Kopan}, {Marsh}, {McCallon}, {Tam}, {Van Dyk}, \&
  {Wheelock}}]{2MASS_survey}
{Skrutskie}, M.~F., {Cutri}, R.~M., {Stiening}, R., {et~al.} 2006, \aj, 131,
  1163

\bibitem[{{Soto} {et~al.}(2007){Soto}, {Rich}, \& {Kuijken}}]{Soto2007}
{Soto}, M., {Rich}, R.~M., \& {Kuijken}, K. 2007, \apjl, 665, L31

\bibitem[{{Spite}(1967)}]{Spite1967}
{Spite}, M. 1967, Annales d'Astrophysique, 30, 211

\bibitem[{{Stetson} \& {Pancino}(2008)}]{Stetson2008}
{Stetson}, P.~B. \& {Pancino}, E. 2008, \pasp, 120, 1332

\bibitem[{{Terndrup} {et~al.}(1998){Terndrup}, {Popowski}, {Gould}, {Rich}, \&
  {Sadler}}]{Terndrup1998}
{Terndrup}, D.~M., {Popowski}, P., {Gould}, A., {Rich}, R.~M., \& {Sadler},
  E.~M. 1998, \aj, 115, 1476

\bibitem[{{Udalski} {et~al.}(1997){Udalski}, {Kubiak}, \&
  {Szymanski}}]{Udalski1997}
{Udalski}, A., {Kubiak}, M., \& {Szymanski}, M. 1997, Acta Astronomica, 47, 319

\bibitem[{{Weidner} \& {Kroupa}(2005)}]{Weidner2005}
{Weidner}, C. \& {Kroupa}, P. 2005, \apj, 625, 754

\bibitem[{{Zhao} {et~al.}(1994){Zhao}, {Spergel}, \& {Rich}}]{Zhao1994}
{Zhao}, H., {Spergel}, D.~N., \& {Rich}, R.~M. 1994, \aj, 108, 2154

\bibitem[{{Zoccali} {et~al.}(2004){Zoccali}, {Barbuy}, {Hill}, {Ortolani},
  {Renzini}, {Bica}, {Momany}, {Pasquini}, {Minniti}, \& {Rich}}]{Zoccali2004}
{Zoccali}, M., {Barbuy}, B., {Hill}, V., {et~al.} 2004, \aap, 423, 507

\bibitem[{{Zoccali} {et~al.}(2008){Zoccali}, {Hill}, {Lecureur}, {Barbuy},
  {Renzini}, {Minniti}, {G{\'o}mez}, \& {Ortolani}}]{Zoccali2008}
{Zoccali}, M., {Hill}, V., {Lecureur}, A., {et~al.} 2008, \aap, 486, 177

\bibitem[{{Zoccali} {et~al.}(2006){Zoccali}, {Lecureur}, {Barbuy}, {Hill},
  {Renzini}, {Minniti}, {Momany}, {G{\'o}mez}, \& {Ortolani}}]{Zoccali2006}
{Zoccali}, M., {Lecureur}, A., {Barbuy}, B., {et~al.} 2006, \aap, 457, L1

\bibitem[{{Zoccali} {et~al.}(2003){Zoccali}, {Renzini}, {Ortolani}, {Greggio},
  {Saviane}, {Cassisi}, {Rejkuba}, {Barbuy}, {Rich}, \& {Bica}}]{Zoccali2003}
{Zoccali}, M., {Renzini}, A., {Ortolani}, S., {et~al.} 2003, \aap, 399, 931

\end{thebibliography}

\end{document}